\documentclass[acmsmall, authorversion]{acmart}

\usepackage{array}
\usepackage{color}
\usepackage{bm}
\usepackage{booktabs}
\newcolumntype{M}[1]{>{\centering\arraybackslash}m{#1}}

\AtBeginDocument{%
  \providecommand\BibTeX{{%
    \normalfont B\kern-0.5em{\scshape i\kern-0.25em b}\kern-0.8em\TeX}}}

\setcopyright{acmcopyright}
\copyrightyear{2021}
\acmYear{2021}
\acmMonth{4}
\acmVolume{15}
\acmNumber{3}
\acmDOI{10.1145/3442200}

\acmJournal{TKDD}

\begin{document}

\title{Neural Networks for Entity Matching: A Survey}

\author{Nils Barlaug}
\orcid{0000-0003-4618-9702}
\affiliation{
    \institution{Cognite}
    \city{Lysaker}
    \country{Norway}
}
\affiliation{
    \institution{NTNU}
    \department{Department of Computer Science}
    \city{Trondheim}
    \country{Norway}
}
\email{nils.barlaug@ntnu.no}

\author{Jon Atle Gulla}
\orcid{0000-0002-9806-7961}
\affiliation{
    \institution{NTNU}
    \department{Department of Computer Science}
    \city{Trondheim}
    \country{Norway}
}
\email{jon.atle.gulla@ntnu.no}

%\renewcommand{\shortauthors}{Nils Barlaug and Jon Atle Gulla}

\begin{abstract}
    Entity matching is the problem of identifying which records refer to the same real-world entity.
    It has been actively researched for decades,
    and a variety of different approaches have been developed.
    Even today, it remains a challenging problem, and there is still generous room for improvement.
    In recent years we have seen new methods based upon deep learning techniques for natural language processing emerge.
    
    In this survey, we present how neural networks have been used for entity matching.
    Specifically, we identify which steps of the entity matching process existing work have targeted using neural networks, and provide an overview of the different techniques used at each step.
    We also discuss contributions from deep learning in entity matching compared to traditional methods,
    and propose a taxonomy of deep neural networks for entity matching.
\end{abstract}



\maketitle

\section{Introduction}
Our world is becoming increasingly digitalized.
While this opens up a number of new, exciting opportunities, it also introduces challenges along the way.
A substantial amount of the value to be harvested from increased digitalization depends on integrating different data sources.
Unfortunately, many of the  existing data sources one wishes to integrate do not share a common frame of reference.
For example, let us say a factory wants to use statistics from equipment maintenance logs to decide which equipment to prioritize for upgrades.
Currently, at this factory, equipment inventory is kept in one system, while maintenance logs are kept in a separate system.
Sadly, these two systems do not refer to equipment in the same way -- i.e., there are no common identifiers or names across the two systems.
While it is possible for a human to identify which maintenance logs belong to which equipment in the inventory system, there is no simple, automatic way to tie the maintenance logs to the inventory records.

Entity matching is the field of research dedicated to solving the problem of identifying which records refer to the same real-world entity.
It is an important data integration task that often arises when data originate from different sources.
The records are usually assumed to either be from two different data sources without duplicates or from the same data source with duplicates.
It is not a new problem.
A group of similar problems has been studied for a long time in a variety of fields under different names (see Section \ref{sec:background}).
Despite having been researched for decades, entity matching remains a challenging problem in practice.
There are several factors that make it difficult in general:
\begin{itemize}
    \item \textbf{
        Poor data quality
    }:
    Real-world data is seldom completely clean, structured, and homogeneous.
    Data originating from manual insertion can contain typos, alternative spellings, or fail to comply with the schema (e.g., mixing first and last name).
    Automatic processes extracting information from unstructured sources might not always be accurate on the scope of attributes (e.g., \texttt{\{firstName: "John Smith", lastName: "Programmer"\}}).
    Furthermore, some data might simply be missing.
    Data in entity matching is often assumed to be structured in records.
    However, it is not unusual that these records are in practice semi-structured because of certain unstructured string attributes -- opening up a world of possible inconsistencies --
    for example, a \texttt{name} attribute (\texttt{"John Smith"}, \texttt{"Smith, John"}, \texttt{"John R. Smith"}, \texttt{"John Richard Smith"}) or an \texttt{adress} attribute.
    In addition, we cannot always expect different data sources to follow the same schema, format, and syntactic conventions.
    \item \textbf{The large number of possible matches}:
    Given $|A|$ records from one data source and $|B| \in \Theta(|A|)$ from another, there are $\Theta(|A|^2)$ possible matches.
    We would normally expect the number of positive matches to be $O(|A|)$.
    This has two important implications.
    First, it is infeasible to explicitly compare all possible pairs for any nontrivial number of records.
    Second, there is an extreme imbalance between positive and negative matches;
    more specifically, there are $\Omega(|A|)$ times as many negative as positive matches.
    The potential for false positives is inherently greater.
    If one wants to use a learning-based approach,
    it can be difficult to label enough positive examples, since they occur in an ocean of negative examples.
    \item \textbf{Dependency on external human knowledge and interaction}:
    The space of potential entity matching problem instances is unbounded and offers great variety.
    While a substantial part of the instances can of course, in theory, be solved automatically,
    in many real-world instances,
    it is either unrealistic or impossible to perform matching as an automatic, isolated process, as the data sources simply do not contain all necessary information.
    Moreover, to perform matching, our solution has to interact with human experts and make use of their knowledge.
    Human interaction is in itself a complex domain.
\end{itemize}

Deep learning has in recent years become an essential part of multiple research fields,
most notably in fields such as computer vision and natural language processing, which are concerned with unstructured data.
Its most prominent advantage over earlier approaches is its ability to learn features instead of relying on carefully handcrafted features \cite{LeCun2015-jz}.
Researchers have already realized the potential advantage of deep learning for entity matching \cite[e.g.,][]{Ebraheem2018-ws, Mudgal2018-cx}.
In this survey, we aim to summarize the work done so far in the use of neural networks for entity matching.

\subsection{Research questions}
One of the challenges of comparing how neural networks are used in entity matching is that published methods often do not address the exact same problem.
They tend to cover somewhat different aspects of entity matching.
With this is in mind, we formulate the following research questions:
\begin{itemize}
    \item How do methods using neural networks for entity matching differ in what they solve, and how do the methods that address the same aspects differ in their approaches?
    \item What benefits and opportunities does deep learning provide for entity matching, and what challenges does it pose?
    \item How can we categorize the different deep neural networks used for entity matching?
\end{itemize}

\subsection{Main contributions}
To answer our research questions, we provide the following main contributions:
\begin{itemize}
    \item We use a reference model of the traditional entity matching process to identify which steps of process that existing work has targeted using neural networks
    and provide an overview of the different techniques that are used for each step.
    \item We discuss the contributions of deep learning to entity matching compared to traditional approaches using a proposed reference model for a deep learning-based entity matching process.
    \item We propose a taxonomy of deep neural networks for entity matching.
    \item We discuss challenges and propose potential future work for deep learning in entity matching understood in the context of our reference entity matching process and deep network taxonomy.
\end{itemize}

\subsection{Outline}
First, as necessary background information, Section~\ref{sec:background} will introduce the problem definition and give a brief introduction to neural networks.
Section~\ref{sec:related-work} mentions related work --- both publications that survey or summarize similar topics and  problems that are similar to entity matching.
We then provide an overview of the surveyed methods using a reference model of the entity matching process as a framework in Section~\ref{sec:process},
before we in Section~\ref{sec:contributions-from-deep-learning} take a step back and discuss contributions from deep learning to entity matching compared to more traditional approaches.
With those contributions in mind, we introduce a taxonomy of deep neural networks for entity matching in Section~\ref{sec:taxonomy}.
Section~\ref{sec:evaluation} provide a brief overview of how evaluation is performed and reported comparative evaluations between deep learning approaches and traditional methods.
Finally, we discuss challenges and opportunities for deep learning in entity matching in Section~\ref{sec:challenges}.

\section{Background}
\label{sec:background}

This section introduces the entity matching problem definition and its many names and variations.
What follows is a brief introduction to neural networks and deep learning and how they are used with text.

\subsection{Problem definition}
Let $A$ and $B$ be two data sources.
$A$ has the attributes $(A_1, A_2, ..., A_n)$, and we denote records as $a = (a_1, a_2, ..., a_n) \in A$.
Similarly, $B$ has the attributes $(B_1, B_2, ..., B_m)$, and we denote records as $b = (b_1, b_2, ..., b_m) \in B$.
A data source is a set of records,
and a record is a tuple following a specific schema of attributes.
An attribute is defined by the intended semantics of its values.
So $A_i = B_j$ if and only if values $a_i$ of $A_i$ are intended to carry the same information as values $b_j$ of $B_j$,
and the specific syntactics of the attribute values are irrelevant.
Attributes can also have metadata (like a name) associated with them, but this does not affect the equality between them.
We call the tuple of attributes $(A_1, A_2, ..., A_n)$ the schema of data source $A$, and correspondingly for $B$.

The goal of entity matching is to find the largest possible binary relation $M \subseteq A \times B$ such that $a$ and $b$ refer to the same entity for all $(a,b) \in M$.
In other words, we would like to find all record pairs across data sources that refer to the same entity.
We define an entity to be something of unique existence\footnote{An entity does not have to be a physical object, but can also be abstract or conceptual -- e.g., a company or an event.}.
Attribute values are often assumed to be strings, but that is not always the case.
It is important to note that the two record sets need not necessarily have the same schema.

Aspects beyond what the surveyed methods cover have been intentionally left out.
For example, we make no matches within each data source, only across the two.
Which is not to say there cannot be duplicates within a data source.
However, in this problem definition, we assume that we are not interested in finding them.
In practice, it is quite common to assume no duplicates within the data sources.
If we are explicitly interested in finding duplicates within a single data source, we can, as will be mentioned below, address duplicates in this formulation of the problem by simply having $A = B$.

In addition, there is also a more subtle assumption in this problem definition:
The record sets $A$ and $B$ are assumed to operate with the same taxonomic granularity.
This is not necessarily always the case.
One data source might refer to households; the other, to individuals,
or two data sources could refer to street-level addresses and postal code areas, respectively.
In many cases, it would still make sense to match records that do not strictly refer to the same entity,
but rather refer to entities with some defined taxonomic closeness.
We leave this out of the definition for simplicity, as it does not affect our analysis of the surveyed methods.

Somewhat ironically, as often pointed out, entity matching itself suffers from the problem of being referenced by many different names,
some referring to the exact same problem, while others are slight variations, specializations, or generalizations.
In addition, the names are not used completely consistently.
Table~\ref{tab:em-names} lists a selection of these names.
We will comment on a few.

\begin{table}[]
    \centering
    \caption{
    Some of the many names that are used for entity matching or similar variations of it.
    }
    \begin{tabular}{l l l}
        Entity matching &
        Entity resolution &
        Record linkage \\
        Data matching &
        Data linkage &
        Reference reconciliation \\
        String matching &
        Approximate string matching &
        Fuzzy matching \\
        Fuzzy join &
        Similarity join &
        Deduplication \\
        Duplicate detection &
        Merge-purge &
        Object identification \\
        Re-identification &
        &
    \end{tabular}
    \label{tab:em-names}
\end{table}

\textit{Entity resolution}, \textit{record linkage}, and \textit{data matching} are frequently used for more or less the same problem as we defined above.
It is not unusual that $A$ and $B$ are assumed to have the same schema --- either because the schemas are, in fact, equal, or because some kind of schema matching has already been performed as a separate step.
Sometimes, fusing the matching pairs to one representation is considered a final step of the problem.
If we also have duplicates within each data source,
it might be necessary to cluster and fuse more than two records at a time.
In this article, we will stick to the more narrow definition laid out above.
\textit{Deduplication} or \textit{duplicate detection} is the problem of identifying which records in the same data source refer to the same entity,
and can be seen as the special case $A = B$.
\textit{String matching} attempts to find strings that refer to the same entity
and can be regarded as the special case $n = m = 1$, if strings are interpreted as single-attribute records.

\subsection{Neural networks and deep learning}

We provide a brief and simplified description of neural networks and deep learning,
followed by a short introduction to how deep learning is used in natural language processing.
A comprehensive introduction to these topics is outside the scope of this survey.
See instead, for example, \citet{Goodfellow2016-my}, from which we will adapt some of our notation in the following paragraphs, for a general introduction to deep learning, and \citet{Goldberg2017-du, Jurafsky2020-uw} for introductions to deep learning for natural language processing.

A \textit{neural network} is a machine learning model.
We wish to approximate some unknown function $f^*(\bm{x})=\bm{y}$ that can map from some interesting input $\bm{x}$ to some desired output $\bm{y}$.
Usually, we will have some examples $D = \{(\bm{x}^{(j)}, \bm{y}^{(j)}) | 1 \leq j \leq m \}$, which are known to be such that $f^*(\bm{x}^{(j)}) \approx \bm{y}^{(j)}$ for all $j$, to help guide us.
To approximate $f^*$ we define a function $f(\bm{x};\bm{\theta})$ parameterized by $\bm{\theta}$,
and then try to learn what $\bm{\theta}$ should be using the examples $D$.
This function $f$ is the neural network.

Even though there are no strict requirements for what constitutes a neural network,
they usually follow a common recipe.
Generally, we let $f$ consist of one or more nested functions $f(\bm{x}) = f_L(f_{L-1}(...f_1(\bm{x})))$.
Each such function $f_l$ would normally be a linear operation, like matrix multiplication, using the parameters $\bm{\theta}$ and then nested by a nonlinear element-wise operation.
For example, $f_l(\bm{x}) = \max(0, W \bm{x} + \bm{b})$, where both $W$ and $\bm{b}$ are part of $\bm{\theta}$ and $\max$ is element-wise.
We call these nested functions \textit{layers},
and $L$ is the \textit{depth} of the network.
When a neural network has several layers (no clear threshold), we call it a deep neural network.

Given a suitable network architecture $f$,
we try to find parameters $\bm{\theta}$ that will make it behave close to the examples $D$.
We first define a loss function $\mathcal{L}(\bm{y}, \hat{\bm{y}})$ quantifying how wrong a prediction $\hat{\bm{y}} = f(\bm{x};\bm{\theta})$ is compared to the correct $\bm{y}$.
Then we randomly initialize $\bm{\theta}$ and perform some variant or descendant of stochastic gradient descent (SGD) with mini-batches:
\begin{equation*}
    \bm{\theta}_{t+1} = \bm{\theta}_t - \alpha \frac{1}{|\widetilde{D}|} \sum_{(\bm{x}, \bm{y}) \in \widetilde{D}} \nabla_{\bm{\theta}_t} \mathcal{L}(\bm{y}, f(\bm{x};\bm{\theta}))
\end{equation*}
where $\alpha$ is the learning rate,
and $\widetilde{D} \subset D$ is a random mini-batch.
The stopping criterion and other details vary between methods.
This procedure is expensive, because it needs to evaluate $f$ and differentiate $\mathcal{L}$ with respect to $\bm{\theta}$.
To make it efficient,
we make sure to choose $|\widetilde{D}| \ll |D|$ and also differentiate with the backpropagation algorithm.
Generally, we can interpret $f$ as a directional acyclic computational graph.
The backpropagation algorithm simply applies dynamic programming using the chain rule over this computational graph.

The real strength of deep learning is its ability to do hierarchical representation learning.
With modern techniques,
multilayered networks are able to learn useful features from relatively unstructured input data \cite{LeCun2015-jz}.
This is especially valuable for data such as images and text, which are notoriously hard to extract good features from with manually crafted procedures.

\subsubsection{Deep learning for natural language processing}
Many state-of-the-art methods for natural language processing are deep learning models \cite[e.g.,][]{Vaswani2017-kk, Devlin2018-sz,Yang2019-hf}.
Central to all these methods is how text is transformed to a numerical format suitable for a neural network.
This is done through embeddings,
which are translations from textual units to a vector space -- traditionally available in a lookup table.
The textual units will usually be characters or words.
An embeddings lookup table can be seen as parameters to the network and be learned together with the rest of the network end-to-end.
That way the network is able to learn good distributed character or word representations for the task at hand.
The words used in a data set are often not unique to that data set, but rather just typical words from some language.
Therefore one can often get a head start by using pretrained word embeddings like word2vec \cite{Mikolov2013-il}, GloVe \cite{Pennington2014-mk} or fastText \cite{Bojanowski2016-dc},
which have been trained on enormous general corpora.
Following a rather recent trend,
large pretrained networks that can produce contextualized word embeddings that take into account the surrounding words are also available \cite{Peters2018-uh, Devlin2018-sz, Radford2019-jl}.

Text is naturally interpreted as a sequence.
It is therefore perhaps not so surprising that neural networks designed for sequences are often used.
One way to model sequences is to use Convolutional Neural Networks (CNNs) --- first popularized by computer vision applications --- which has received considerable attention within the natural language processing community \cite{Kim2014-cx,Conneau2016-pl}.
However, a more prominent sequence model has been Recurrent Neural Networks (RNN) \cite{Elman1990-px} and their variants.
RNNs are constructed by repeating the same layer multiple times.
Each layer takes both the output from the previous layer as well as some part of an input sequence.
So assuming the input to be a sequence $\bm{x_1}, ..., \bm{x_L}$,
we nest layers recursively as $\bm{h_l} = f_l(\bm{h_{l-1}}, \bm{x_l}; \bm{\theta})$,
where $\bm{h}_l$ is called the hidden state.
Layers share the same parameters,
and the number of layers can therefore be dynamically adjusted to the length of the input sequence.
The last hidden state will, in theory, contain information about the whole input sequence.
Additional layers can be appended to further process this feature vector and produce some desired output.
Output sequences can be generated in a number of ways by setting the initial hidden state and then extracting the hidden state from different layers.
RNNs themselves consist of a (dynamic) number of layers,
but it is also possible to nest several RNNs.
We then get what is called stacked RNNs.

RNNs are relatively deep networks and are therefore prone to what is called vanishing gradients.
The gradients from the early layers become so small that they are ineffective in gradient descent.
In other words, the first parts of the input sequence have too little influence over the end result.
Therefore, variants of RNNs such as Long Short-Term Memory (LSTM) \cite{Hochreiter1997-qz} and Gated Recurrent Units (GRU) \cite{Cho2014-yy} are often used in practice.
They make sure that hidden states are more easily able to flow through the subsequent layers undisturbed,
so that gradients will remain strong when backpropagated through many layers.
Despite this improvement, the networks will still tend to be influenced more by the end of the input sequence then the beginning.
It has become quite common to have bidirectional RNNs \cite{Schuster1997-qf, Graves2005-su},
which can be seen as combining two RNNs, where one of them processes the input sequence backward.

Another popular way to face the issue of skew in influence for sequences is to use attention mechanisms \cite{Bahdanau2014-mk}.
The idea is to let the network itself choose what parts of the input to focus on,
potentially for several iterations.
This is typically achieved in a network by producing some normalized attention vector that is multiplied with the vector of interest.

\begin{figure}
    \centering
    \includegraphics{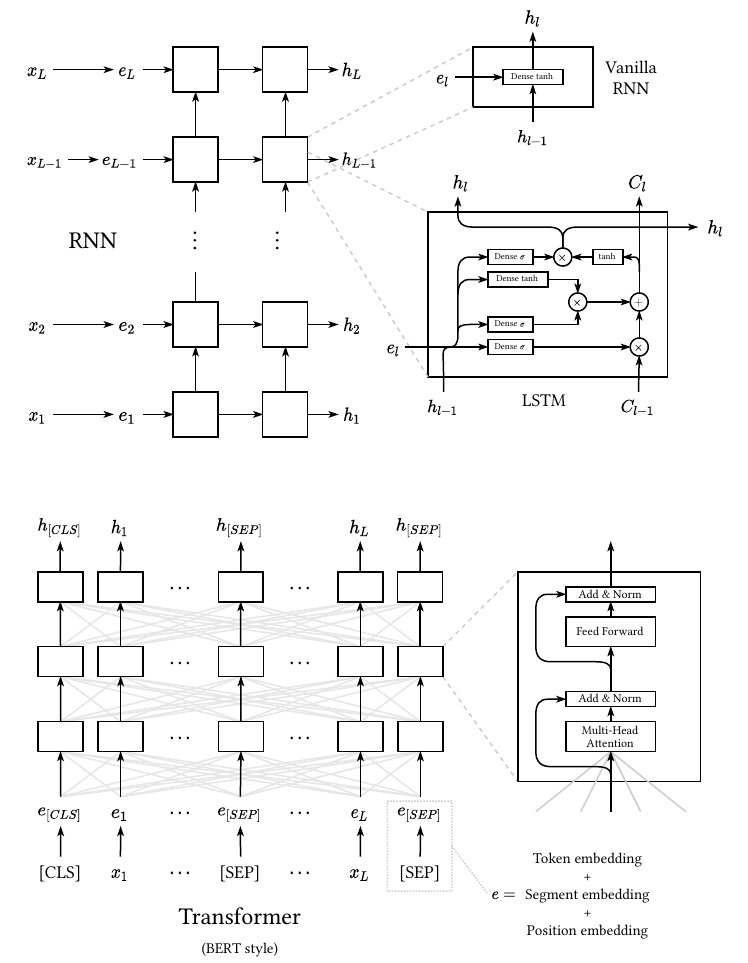}
    \caption{
        Illustration of the architecture for a two-stack uni-directional RNN encoder and a three-layer BERT-style \cite{Devlin2018-sz} encoder for natural language processing.
        Let $(x_1, x_2, \dots, x_l, \dots, x_L)$ be the input sequence, and $e_l$ be an embedding for $x_l$.
        Both a standard RNN and LSTM block are illustrated for the RNN architecture.
        Notice the additional \textit{context} state $C_l$ for LSTM, which can more easily carry gradients.
        Inspired by illustrations in \cite{Olah2015-rj,Vaswani2017-kk,Devlin2018-sz}.
    }
    \label{fig:rnn-transformer}
\end{figure}

While initially used as an enhancement to RNNs,
networks based almost solely on attention \cite{Vaswani2017-kk} have recently started to proliferate \cite{Devlin2018-sz, Radford2019-jl, Yang2019-hf}
and are currently considered state-of-the-art for many, if not most, natural language processing tasks.
We call these Transformer-based networks --- as originally named by \cite{Vaswani2017-kk} that targeted machine translation.
In contrast to RNN-based networks, they are not sequential with respect to the input sequence.
See Figure~\ref{fig:rnn-transformer} for an illustration of an RNN and Transformer encoder.
This makes them more parallel, which again makes it easier to leverage modern, highly parallel hardware.
In addition, one avoids prohibitive deep networks (due to vanishing gradients) for long input sequences. 
Each layer performs self-attention over the whole input sequence, effectively removing the long paths between cells of RNNs that makes it so hard to learn long-range dependencies.
Since transformer networks are architecturally agnostic to the input sequence order,
they are instead fed positional information through the input as positional embeddings.

One particular influential recent trend has been the ability to leverage huge pretrained models that have been trained unsupervised for language modeling on massive text corpora \cite{Peters2018-uh, Devlin2018-sz, Radford2019-jl} --- similar to what the computer vision community has done for a while.
They produce contextualized word embeddings that take into account the surrounding words.
The embeddings can be used as a much more powerful variant of the classical word embeddings,
but as popularized by BERT \cite{Devlin2018-sz}, one can also fine-tune the network to the task at hand.
Take BERT as an example.
It is pretrained jointly on masked language modeling and next sentence prediction.
Input during training is a special [CLS] token first, then the two sentences terminated by a special [SEP] token each.
The [CLS] tokens output from the network is used to do the next sentence classification.
Each token's embedding is augmented with a positional embedding and a segment embedding indicating which sentence it belongs to.
This setup makes the network suitable for fine-tuning on both sequence labeling tasks as well as pair labeling tasks (such as question answering or entity matching).

\section{Related work}
\label{sec:related-work}

\subsection{Other surveys and extensive overviews}
Given entity matching's long history, there is no surprise that it has been surveyed before in various ways,
covering entity matching as a whole and more narrow aspects.

First, there are several books that provide an overview.
\citet{Christen2012-xr} is a dedicated and comprehensive source on entity matching,
\citet{Naumann2010-rb} specifically cover the slightly more specialized problem of duplicate detection, and
\citet{Batini2006-su,Talburt2011-zf,Doan2012-gj} all introduce entity matching in the context of data quality and integration.
Second, the workshop tutorials by \citet{Getoor2012-hx,Stefanidis2014-qk} serve as introductory summaries.
Third, \citet{Elmagarmid2007-ow} present a literature analysis.

Other sources cover more narrow aspects of entity matching -- such as specific techniques or subtasks.
Quite early on, statisticians dominated the field of entity matching.
Probabilistic methods were developed by \citet{Newcombe1959-ov} and given a solid theoretical framework by \citet{Fellegi1969-nc}.
These probabilistic methods are summarized by \citet{Winkler1995-xq,Winkler2006-rc, Herzog2007-ha}.
Blocking, which is surveyed by \citet{Papadakis2016-ka,Christen2012-wi,Papadakis2019-pb}, is considered an important subtask of entity matching, meant to tackle the quadratic complexity of potential matches.
\citet{Christophides2019-de} specifically review entity matching techniques in the context of big data.
There has been an uptick in interest in both machine learning and crowdsourcing as a solution to entity matching in recent years.
As part of a larger survey on crowdsourced data management, \citet{Li2016-hb} cover crowdsourced entity matching.
\citet{Lu2019-xf} summarize the use of machine learning, while \citet{Gurajada2019-yk} present an overview of crowdsourcing, active learning, and deep learning for entity matching.

While earlier works mention or cover neural networks for entity matching to various degrees,
we are to the best of our knowledge the first to present a dedicated, complete, and up-to-date survey.

\subsection{Related problems}
Entity matching can be seen as part of a larger group of tasks with roots in natural language processing that solve similar, but distinct, matching problems.
Interestingly, but perhaps not surprisingly, deep learning-based methods have become state-of-the-art in all these tasks.
We will briefly mention some of the most prominent ones.
\begin{itemize}
    \item \textbf{Coreference resolution}:
    Given a text, find all mentions of entities and determine which mentions corefer.
    Two entity mentions corefer if they refer to the same entity \cite{Jurafsky2008-ii}.
    In contrast, entity matching is concerned with more structured data with clearly distinct units of data (records).
    Importantly, entity matching does not have to take into account a larger textual context,
    which is necessary in coreference resolution to find coreferring mentions across multiple sentences.
    State-of-the-art methods are able to perform the whole task end-to-end using a deep network without detecting and disambiguating mentions in two separate steps \cite{Lee2017-do, Joshi2019-cn}.
    \item \textbf{Entity alignment}:
    Given two knowledge bases, find which entries across the two that refer to the same entity.
    Knowledge bases, in contrast to record sets in entity matching, have relations between entries.
    Leveraging these relations are central to the task.
    The way most neural-based methods do this is by producing so-called knowledge graph embeddings \cite{Zhu2017-sa, Sun2018-qa, Chen2018-ud},
    embeddings of entries which incorporate information about their relationship to other entries.
    
    As a slightly specialized variant, user identity linkage is the problem of identifying which users across two social networks are the same \cite{Zhong2018-qk}.
    \item \textbf{Entity linking}:
    Given a text, find all mentions of entities and link them to entries in a knowledge base.
    One example of a heavily used knowledge base would be Wikipedia.
    In some ways,
    one can see entity linking as a hybrid between coreference resolution and entity alignment,
    and it differs from entity matching in the same ways.
    Neural-based methods are considered state-of-the-art \cite{Raiman2018-sp, Kolitsas2018-jl}.
    \item \textbf{Paraphrase identification}:
    Given two texts, determine if they are semantically equivalent -- i.e., if they carry the same meaning.
    This can be be seen as a generalization of string matching, if one interprets strings referring to the same entity as implicating that the strings are also semantically similar.
    Nonetheless,
    we still consider figuring out which texts convey the same meaning in general to be a distinct problem from entity matching.
    First, entity matching deals with more structured data.
    Second and most importantly, in entity matching, all records refer to an entity, and we are only concerned with which specific real-world entity a record is referring to.
    Any excess meaning carried by a record does not impact matching.
    
    Finally, there is also semantic textual similarity and textual entailment, which are closely related to paraphrase identification.
    Semantic textual similarity is concerned with the degree of how semantically similar two texts are,
    while textual entailment is about finding out whether one text semantically entails, contradicts, or is neutral to a second text.
    Additionally, in the case of multiple choice, question answering can also be seen as a matching problem.
    
    State-of-the-art for most of these matching problems rely on rather generic, but powerful, language understanding models \cite{Yang2019-hf, Devlin2018-sz}.
\end{itemize}

In a broader sense,
similar problems are also studied in the context of information retrieval \cite{Mitra2017-wj}.
Neural networks not only provide effective techniques for retrieving unstructured text but also for data formats that have traditionally been less accessible such as images \cite{Lin2015-hy} --- and even across modals \cite{Wu2019-xd}.

\section{The entity matching process}
\label{sec:process}

\begin{figure}
    \centering
    \includegraphics{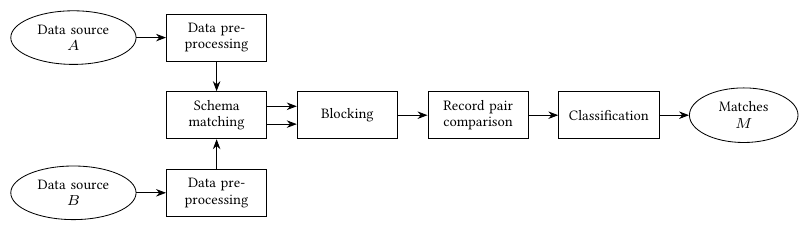}
    \caption{
    Illustration of the reference model for a traditional entity matching process and its five steps.
    Human-in-the-loop aspects are not considered.
    }
    \label{fig:process}
\end{figure}

Traditionally, entity matching is often thought of as a process consisting of multiple steps,
even though there is no generally agreed upon list of specific steps.
It is useful to compare methods in light of how they relate to this abstract process.
To this end, we introduce a high-level reference model describing the entity matching process as five distinct steps.
These steps can also be viewed as a chain of the subtasks or subproblems that make up entity matching.
Inspired by processes and figures such as those in \cite{Naumann2010-rb, Christen2012-xr, Ebraheem2018-ws, Christophides2019-de, Gurajada2019-yk}, Figure~\ref{fig:process} depicts this reference model of the traditional entity matching process.
We will use the model to frame the discussion of different methods using neural networks.

The process adheres to the problem definition introduced above.
It assumes two data sources as input.
In theory, it could be generalized to multiple sources, but this is seldom done in the literature.
A single source, as previously mentioned, can simply be seen as a special case.
At the end of the process, the result is simply matches.
Since this is an abstract process extracted from the literature,
it is not necessarily followed step by step.
The order might not be completely strict, and steps might be intermingled or skipped -- as will be clear when we look at specific methods.

We also note that this process is machine-oriented
and does not highlight any iterative human interactions or feedback loops.
Significant research has gone into both crowdsourcing \cite{Vesdapunt2014-ig, Wang2012-lr, Whang2013-gu, Govind2018-ke} and active learning \cite{Kasai2019-ii, Arasu2010-ex, Qian2017-ec}.
Interestingly, \citet{Kasai2019-ii} use a deep neural network in their active learning approach.
Such human-in-the-loop factors are often crucial for entity matching in practice \cite{Doan2017-pi}.
We do not consider our proposed process to be in conflict with these aspects,
but rather mostly orthogonal.
Empirically, based on the surveyed methods, we do not find neural networks to be very tightly coupled to any human-in-the-loop techniques.
We therefore focus on the machine-oriented aspects.

\subsubsection{Data preprocessing}
The first step in the process is data preprocessing, which is usually a crucial step in many data integration tasks \cite{Doan2012-gj}.
The goal is to get both data sources into consistent and similar formats better suited for downstream tasks.
Typical transformations may involve removing excess punctuation, lowercasing all letters, normalizing values, and tokenizing.
Sometimes, one might also view this step as feature extraction, where records are transformed to a feature space.
Preprocessing is, of course, very dependent on the domain and the specific data sources.

\subsubsection{Schema matching}
After preprocessing we perform schema matching,
where the objective is to find out which attributes should be compared to one another,
essentially identifying semantically related attributes.
This will enable downstream steps to compare records across the two sources.
Even though schema matching is often considered a separate problem to be solved before performing entity matching \cite[e.g.,][]{Christen2012-xr},
we choose to include it both because deep learning-based methods have the potential to perform it jointly with other steps (as a surveyed method shows \cite{Nie2019-tw}) and because it is frequently an unavoidable problem in real-world use cases for entity matching.

In practice,
this step is often performed manually as part of the preprocessing step,
simply making sure to transform both data sources into the same schema format.
Traditional techniques for schema matching span a wide range of solutions.
They can use both schema metadata and actual attribute values.
Some are supervised learning methods, while others are unsupervised.
Importantly, most of them are completely independent of downstream tasks in the process,
though most techniques are actually not developed specifically for the purpose of entity matching.
For more in-depth coverage of schema matching, see \citet{Rahm2001-xl, Bellahsene2011-hh, Doan2012-gj}.

\subsubsection{Blocking}
Since the number of potential matches grows quadratically,
it is common to pick out candidate pairs $C \subseteq A \times B$ in a separate step before any records are compared directly.
We call this step \textit{blocking}, and the goal is to bring down the number of potential matches $|C| \ll |A \times B|$ to a level where record-to-record comparison is feasible.
Note that in the literature, blocking is sometimes used as a name for only one of the possible strategies for avoiding the quadratic complexity \cite[e.g.,][]{Christophides2019-de}.
For simplicity, we refer to any effort to make record comparison feasible as blocking.

One can think of the blocking step as doing implicit comparison of records,
while the comparison step described below is doing explicit comparison between pairs of records.
There is often no way around performing at least some explicit pairwise comparison, since implicit comparison cannot offer the necessary precision.
In certain cases, when the comparison lends itself to indexing, it is possible to fuse record pair comparison and blocking into one step.
Usually, however, the explicit comparison is nontrivial, infeasible, or impossible to speed up by indexing --
necessitating a need to prune away obvious nonmatches in a separate blocking step.
This is possible because implicit comparison will typically have lower precision, but can be done more efficiently.
At the same time, explicit comparison will typically have higher precision but has inherent quadratic complexity.
By constructing a high-recall implicit comparison step to filter away obvious nonmatches first,
we can make it feasible to use a more powerful high-precision explicit comparison afterward.

Typical techniques are based on hashing, sorting, or various ways of indexing.
Some work completely independent from the downstream steps,
while others are more coupled with the record pair comparison and classification steps.
For example,
if matches are decided based on thresholds of string similarity measures,
it is often possible to specifically index attribute values to prune away according to that criteria \cite{Yu2016-fo}.
Most techniques rely heavily on syntactic similarity,
including those based on supervised machine learning.
See \citet{Christen2012-wi,Christophides2019-de} for extensive reviews on blocking techniques.
In practice,
it is not uncommon that blocking involves quite a bit of manual feature selection,
picking out which attributes should be used and which technique to apply.

\subsubsection{Record pair comparison}
When the number of candidate pairs $|C|$ has been reduced to a manageable amount, we can compare individual records $(a,b) \in C$.
The pairwise comparison results in a similarity vector $S$, consisting of one or more numerical values indicating how similar or dissimilar the two records are.

Traditionally,
such comparisons have mostly been made using string similarity measures.
These measures typically quantify a very specific syntactic similarity,
and therefore differ in what clues for matching strings they are able to pick up.
Some are, for example, good at adjusting for spelling errors or OCR errors.
String similarity measures have been extensively covered before \cite{Elmagarmid2007-ow, Christen2012-xr, Lu2019-xf}.
It is possible to incorporate domain knowledge in a string similarity measure to also perform semantic comparison instead of just syntactic \cite{Arasu2008-mc, Shang2016-ut},
but it is less common and introduces the extra challenge of acquiring such materialized domain knowledge.

String similarity measures are made to compare two strings and cannot be directly applied to a pair of records.
Normally, one will compare those attributes which were found to be semantically similar in the schema matching step,
 thereby getting multiple measurements to include in $S$.
Also, since the similarity measures are almost always static and only cover a narrow syntactic similarity,
one can use multiple measures and offload the job of figuring out which ones to emphasize to the downstream classification step.

\subsubsection{Classification}
Lastly, the objective of the classification step is to classify each candidate pair as either match or nonmatch based on the similarity vector.
In cases where $|S|=1$, simple thresholding might be enough,
while when $|S|>1$, one needs more elaborate solutions.

Early efforts in entity matching were focused on unsupervised probabilistic methods for doing classification.
Initially developed by \citet{Newcombe1959-ov} and later formalized by \citet{Fellegi1969-nc},
the idea is that, given certain assumptions,
one can calculate the optimal matching choice according to some bounds on false positives and negatives.
It can be seen as very close to a näive Bayes classifier,
classifying record pairs as either match, nonmatch, or uncertain -- where the uncertain matches must go through manual review.
The motivation is that common attribute values that agree (for example, a very common first name) are less significant than rare attribute values that agree.
See \citet{Herzog2007-ha} for a complete introduction to probabilistic approaches.

Recently, methods based on rules or machine learning have been more prominent.
Rules are predicates over the similarity vector $S$ that flag them as match or nonmatch.
They can be constructed manually, making them a powerful and highly explainable way of explicitly incorporating domain knowledge into the classification.
Manually constructing rules requires a lot of expert labor,
so significant work has been put into automatically learning rules from examples.
Other efforts in leveraging learning have used off-the-shelf classification models such as decision trees and support vector machines.
These machine learning models are then trained on examples of $S$ for which it is known if they represent a matching or nonmatching record pair.
Both rule-based and machine learning approaches are covered extensively in the literature \cite{Elmagarmid2007-ow, Doan2012-gj, Christen2012-xr}.

\subsubsection{Outline}
Table~\ref{tab:method-process-matrix} lists, to the best of our knowledge, all methods that use neural networks for entity matching and which steps of the process they tackle using neural networks.
We will in the subsequent subsections take a closer look at each step
and see how different methods use neural networks to handle them.

\begin{table}
    \centering
    \caption{
        Overview of which steps of the entity matching process reference model different methods tackle with neural networks.
    }
    \small
    \begin{tabular}{l|M{17mm}|M{13mm}|M{13mm}|M{17mm}|M{17mm}}
        Method & Data \newline preprocessing & Schema matching & Blocking & Record pair \newline comparison & Classification \\
        \hline
        SEMINT \cite{Li1994-sd,Li2000-le}&           & $\bullet$ &           &           &           \\ \hline
        SMDD \cite{You_Li2005-jw}        &           & $\bullet$ &           &           &           \\ \hline
        \citet{Nin2006-ld}               &           & $\bullet$ &           & $\sim$    &           \\ \hline
        \citet{Pixton2006-ko}            &           &           &           &           & $\bullet$ \\ \hline
        \citet{Wilson2011-qe}            &           &           &           &           & $\bullet$ \\ \hline
        \citet{Tran2014-ww}              &           &           &           &           & $\bullet$ \\ \hline
        NNSM \cite{Zhang2014-bu}         &           & $\bullet$ &           &           &           \\ \hline
        \citet{Gottapu2016-ho}           & $\bullet$ &           &           &           & $\bullet$ \\ \hline
        \citet{Reyes-Galaviz2017-ek}.    &           &           &           &           & $\bullet$ \\ \hline
        % \hline
        
        \citet{Kooli2018-sr}             & $\bullet$ &           &           & $\bullet$ & $\bullet$ \\ \hline
        DeepMatcher \cite{Mudgal2018-cx} & $\bullet$ &           &           & $\bullet$ & $\bullet$ \\ \hline
        \citet{Wolcott2018-zb}           & $\bullet$ &           &           & $\bullet$ &           \\ \hline
        DeepER \cite{Ebraheem2018-ws}    & $\bullet$ &           & $\bullet$ & $\bullet$ & $\sim$    \\ \hline
        MPM \cite{Fu2019-ko}             & $\bullet$ &           &           & $\bullet$ & $\bullet$ \\ \hline
        \citet{Kasai2019-ii}             & $\bullet$ &           &           & $\bullet$ & $\bullet$ \\ \hline
        Seq2SeqMatcher \cite{Nie2019-tw} & $\bullet$ & $\bullet$ &           & $\bullet$ & $\bullet$ \\ \hline
        \citet{Nozaki2019-sg}            & $\bullet$ & $\bullet$ &           &           &           \\ \hline
        AutoBlock \cite{Zhang2019-or}    & $\bullet$ &           & $\bullet$ &           &           \\ \hline 
        Hi-EM \cite{Zhao2019-yv}         & $\bullet$ &           &           & $\bullet$ & $\bullet$ \\ \hline
        \citet{Brunner2020-tn}           & $\bullet$ & $\bullet$ &           & $\bullet$ & $\bullet$ \\ \hline
        Ditto \cite{Li2020-jp}           & $\bullet$ & $\bullet$ &           & $\bullet$ & $\bullet$
    \end{tabular}
    \label{tab:method-process-matrix}
\end{table}

\subsection{Data preprocessing}
\label{sec:data-preprocessing}

Deep neural networks are good at doing representation learning.
As we will see, they can therefore effectively learn to do some of the data preprocessing we would traditionally do manually.
When we explore how the different methods do this,
we will focus on two aspects:
How embeddings are used to get records in a suitable input format, and how the networks' hierarchical representations are structured.

\begin{table}
    \centering
    \caption{
    Overview of how the surveyed methods use embeddings,
    specifically at what granularity, if they use pretrained embeddings, and whether they fine-tune embeddings.
    Surveyed methods not using embeddings at all are left out.
    $^+$Other options were tried, but this was found to be most preferential.
    $^*$The method uses pretrained embeddings for the attribute value text,
    but standard embeddings trained from scratch for attribute labels.
    }
    \small
    \begin{tabular}{llm{15mm}m{15mm}}
        \toprule
        Method & Embedding granularity & Pretrained \newline embeddings & Fine-tuned \newline embeddings \\ \midrule
        \citet{Gottapu2016-ho}           & Word      & No & -   \\ % \hline
        
        \citet{Kooli2018-sr}             & Word \& Character N-gram     & Yes     & No      \\ % \hline
        DeepMatcher \cite{Mudgal2018-cx} & Word \& Character N-gram$^+$ & Yes$^+$ & No      \\ % \hline
        \citet{Wolcott2018-zb}           & Character                    & No      & -       \\ % \hline
        DeepER \cite{Ebraheem2018-ws}    & Word                         & Yes     & Yes$^+$ \\ % \hline
        MPM \cite{Fu2019-ko}             & Word \& Character N-gram     & Yes     & No      \\ % \hline
        \citet{Kasai2019-ii}             & Word \& Character N-gram     & Yes     & No      \\ % \hline
        Seq2SeqMatcher \cite{Nie2019-tw} & Word \& Character N-gram     & Both$^*$     & No \\ % \hline
        \citet{Nozaki2019-sg}            & Word                         & Yes     & No      \\ % \hline
        AutoBlock \cite{Zhang2019-or}    & Word \& Character N-gram     & Yes     & No      \\ % \hline 
        Hi-EM \cite{Zhao2019-yv}         & Character                    & No      & -       \\
        \citet{Brunner2020-tn}           & Character N-gram             & Yes     & Yes     \\
        Ditto \cite{Li2020-jp}           & Character N-gram             & Yes     & Yes     \\
        \bottomrule
    \end{tabular}
    \label{tab:feature-extraction}
\end{table}

\subsubsection{Embeddings}
\label{sec:embeddings}
Neural networks alone only work with numerical data,
so an important enabling factor in letting networks learn representations is how textual records are transformed into a numerical format.
In practice, this is done using embeddings, as explained in Section~\ref{sec:background}.
Note that while some embedding models, like GloVe \cite{Pennington2014-mk}, are not neural networks,
we still consider them a crucial component for neural networks and how they are able to replace manual feature extraction (see Section~\ref{sec:contributions-from-deep-learning}).
They perform and enable powerful representation learning on text.
Other embedding models, like word2vec, can be seen as a simple neural network.
Even though the embeddings are later used in a lookup table, they were trained using this simple network.
One interesting use of word2vec is that of \citet{Nozaki2019-sg}.
They do not use the word embeddings as input to a neural network,
but use them as is in a simple aggregation and comparison scheme to do schema matching (details in Section \ref{sec:schema-matching-attribute-embeddings}).

\paragraph{Granularity}
Embeddings can be used at different granularities, usually at word- or character-level.
The second column of Table~\ref{tab:feature-extraction} shows which methods use which granularity.
Word embeddings significantly reduce the length of the sequences to be processed compared to character embeddings
but come at the expense of increasing the number of unique values that have to be represented by many orders of magnitude.
This often makes solutions relying on word embeddings more vulnerable to out-of-vocabulary (OOV) words --
i.e., words that were not present in the training data.
Word-based embedding models usually handle unknown words by assigning the same embedding to all unknown words, making no distinction between them.
When embeddings are pretrained on large general corpora (as will be discussed next),
but the data sources at hand contain domain-specific words that are otherwise rare,
they will naturally tend to be less useful.
In addition, the data sources at hand might have low data quality and contain typos or small spelling variations that are not common in the training data -- thus effectively making those words out-of-vocabulary.
Motivated by these concerns, the majority of the methods use fastText \cite{Bojanowski2016-dc} embeddings (or similarly, Wordpiece/SentencePiece/Byte-Pair-Encoding \cite{Schuster1997-qf,Kudo2018-wn,Sennrich2016-ad} for the transformer networks).
FastText combines embeddings for both the word itself and all character N-grams of certain lengths,
often making it possible to find a suitable representation for an OOV word, since the word most likely has known N-grams.
Using N-grams in this way is basically a way of approximately incorporating a morpheme granularity level to word-level embeddings \cite{Bojanowski2016-dc}.

Does the choice of embeddings matter?
\citet{Mudgal2018-cx} compare fastText to (the purely word-based) GloVe
and find fastText to have an edge when the data sources contain domain-specific words that are OOV and otherwise comparable.
\citet{Ebraheem2018-ws}, meanwhile, compare fastText without N-grams, GloVe, and word2vec \cite{Mikolov2013-il}, reaching the conclusion that there is no significant difference.
The combined results might indicate that the embedding granularity is more important than which particular embedding is used.

\paragraph{Pretraining}
One of the benefits of popular word embeddings models like word2vec, GloVe, and fastText is that you can get pretrained embeddings.
They have been trained on enormous general corpora, have vocabularies of significant size, and are often available for different languages.
Pretrained character embeddings are not nearly as common,
though \citet{Zhao2019-yv} are, in essence, pretraining the entity matching model on large amounts of training data and can in a way be thought of having pretrained character embeddings.
The third column in Table~\ref{tab:feature-extraction} shows which methods use pretrained embeddings.
Using pretrained embeddings is essentially a way of doing transfer learning for feature extraction.
Since embeddings can be trained unsupervised,
there is generally a substantial amount of training data available.
This can be very helpful if it manages to reduce the necessary amount of labeled training data for the downstream entity matching task at hand.
\citet{Mudgal2018-cx} found that learning embeddings from scratch instead of using pretrained embeddings can be favorable to highly specialized data sources,
while for other data sources,
pretrained embeddings either outperformed or were comparable to learning from scratch.

\paragraph{Fine-tuning}
Even when embeddings have been pretrained on some large text collection,
one still has the opportunity to continue adjusting them when doing the task-specific training together with all other weights.
We refer to this as \textit{fine-tuning} the embeddings -- the opposite of \textit{freezing} them.
The fourth column in Table~\ref{tab:feature-extraction} shows which methods fine-tune their embeddings, which currently is only \citet{Ebraheem2018-ws}.
They found fine-tuning to help on hard data sets --- i.e., those that are very challenging or impossible to get close to perfect $F_1$ score on.

\subsubsection{Representation levels}
\label{sec:representation-levels}
Embeddings offer neural networks an initial mapping from the actual input to a suitable numeric representation.
But as mentioned earlier, the strength of deep learning's use of neural networks is really its ability to do hierarchical representation learning,
which is achieved using multiple layers, learning increasingly abstract features \cite{LeCun2015-jz}.
The first layers of deep networks will typically be designed to enable building a good representation of the input, and then let the last layers focus on producing the desired output.
It is nontrivial to figure out what each layer actually learns.
When we compare the surveyed methods,
we instead focus on explicit levels of representation.

\begin{table}
    \centering
    \caption{
    Overview of which explicit representation levels the surveyed methods make use of and which kinds of network layers are used to build representation from the level below.
    Methods upper half use independent record representations,
    and those in the bottom half use interdependent record representations.
    Self-attention is any attention mechanism that only uses elements within the same record,
    while cross-attention refers to any attention mechanism looking across two records.
    $^+$Other options were explored, but the representational power was similar or lower.
    $^*$Multiple options in use at the same time.
    }
    \footnotesize
    \begin{tabular}{lm{15mm}m{20mm}m{25mm}m{25mm}}
        \toprule
        Method & Character & Word & Attribute & Record \\ \midrule % \hline
        \citet{Gottapu2016-ho}           &           & Standard \newline embeddings &           & 1 Convolutional       \\ \addlinespace[0.5em] %\hline
        
        \citet{Wolcott2018-zb}           & Standard \newline embeddings & & & 1 BiLSTM, 2 FC \\ \addlinespace[0.5em] %\hline
        DeepER \cite{Ebraheem2018-ws}    &           & GloVe$^+$ &           & 1 BiLSTM$^+$ \\ \addlinespace[0.5em] %\hline
        MPM \cite{Fu2019-ko}             & $^*$      & fastText$^*$ & $^*$ &           \\ \addlinespace[0.5em] %\hline
        \citet{Kasai2019-ii}             &           & fastText & 1 BiGRU &           \\ \addlinespace[0.5em] %\hline
        \citet{Nozaki2019-sg}            &           & word2vec & Sum, average &           \\ \addlinespace[0.5em] %\hline
        AutoBlock \cite{Zhang2019-or}    &           & fastText & 1 BiLSTM$^+$, \newline self-attentation & Weighted sum \\ \midrule %\hline 
        %\hline
        DeepMatcher \cite{Mudgal2018-cx} &           & fastText$^+$ & Cross-attention, \newline 1 BiGRU$^+$ & \\ \addlinespace[0.5em] %\hline
        Seq2SeqMatcher \cite{Nie2019-tw} &           & Standard \newline embeddings \& \newline fastText &           & Cross-attention    \\ \addlinespace[0.5em] %\hline
         
        Hi-EM \cite{Zhao2019-yv}         & Standard \newline embeddings & 1 BiGRU, \newline cross-attention, \newline self-attention & 1 BiGRU, \newline cross-attention, \newline self-attention & Concatenation   \\ \addlinespace[0.5em]
        \citet{Brunner2020-tn}           & \multicolumn{2}{m{35mm}}{Byte-pair encoding$^+$, \newline 12 transformer layers \newline (self- and cross-attention)$^+$} & & \\ \addlinespace[0.5em]
        Ditto \cite{Li2020-jp}           & \multicolumn{2}{m{35mm}}{Byte-pair encoding$^+$, \newline 12 transformer layers \newline (self- and cross-attention)$^+$} & & \\
        \bottomrule
    \end{tabular}
    \label{tab:representation-levels}
\end{table}

\begin{figure}
    \centering
    \includegraphics[width=\textwidth]{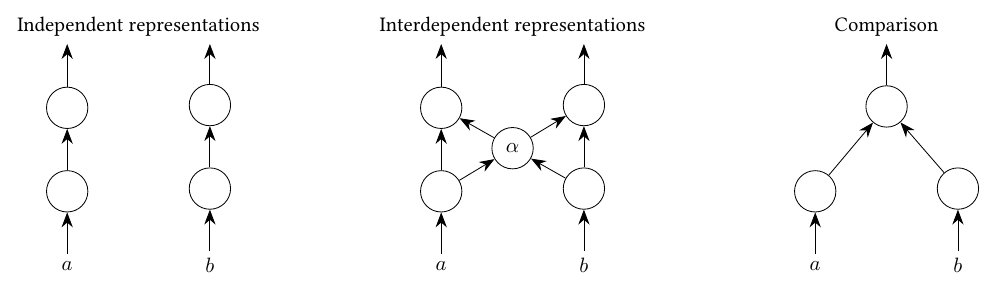}
    \caption{
        Illustration of what is considered a vector representation (independent or interdependent) and what is considered a comparison in terms of computational graphs.
        Here, $a$ and $b$ are records and $\alpha$ is an attention mechanism.
    }
    \label{fig:representation-comparison}
\end{figure}

Table~\ref{tab:representation-levels} highlight how each method does representation learning by listing which explicit representation levels are used,
and which techniques are used to build representation from the level below,
where the explicit level units are character, word, attribute, and record.
It is important to note that the table does not reflect the entire neural network of each method,
but rather only the beginning layers that are to be considered as the feature extraction part of the network.
We consider a representation level as used if you can simply pick out a vector representation for units of that level after some layer.
A vector is considered to represent a unit if its calculations only rely on that unit or other input through an attention mechanism.
Importantly,
a vector that relies on two records through something else than an attention mechanism is not considered a representation,
but rather a comparison (see Section~\ref{sec:comparison}).
Figure~\ref{fig:representation-comparison} illustrates the difference in terms of computational graphs.
Of course, with neural networks, the actual line between the initial feature extraction part and the rest is an artificial one and not necessarily indicative of how the networks actually learn and work.
But they do reflect design decisions to a certain degree and help us compare them in that regard.

We see each method's first layer is (unsurprisingly) an embedding,
providing initial character or word vectors.
Some use a specific embedding model, like fastText, while others just use standard lookup table embeddings that they train themselves.
Next, we note the popularity of RNN-based models among the methods,
which is in line with the widespread use of such sequence-aware models in natural language processing \cite[e.g.,][]{Graves2013-uh, Sutskever2014-ya, Lample2016-xo}.
An interesting case is that of MPM \cite{Fu2019-ko}, which actually combines two versions of DeepMatcher \cite{Mudgal2018-cx} as well as classical similarity measures in its architecture.

The methods can be naturally divided into two categories when it comes to representation learning:
independent or interdependent representation.
If the highest representation level relies on a record pair instead of a single record,
we say it is an interdependent representation.
Otherwise, it is an independent representation.
See again Figure~\ref{fig:representation-comparison} for an illustration.
The methods in Table~\ref{tab:representation-levels} have independent and interdependent representation at the top and bottom, respectively.
Interdependent representations are, in essence, a way to incorporate record pair comparison into the feature extraction.
They have the benefit of being able to adapt based on what they will be compared to,
while independent representations have the benefit of not relying on record pairs to be computed.
The latter will be important when we discuss blocking in Section~\ref{sec:blocking}.

\paragraph{Independent representation}

There is significant variation among the methods with independent representation.
\citet{Kooli2018-sr} and \citet{Nozaki2019-sg} mostly rely on word embeddings for the feature extraction part of the network.
\citeauthor{Kooli2018-sr} simply concatenate them before the next layers do comparison,
and \citeauthor{Nozaki2019-sg} aggregate them through summation and averaging.
\citet{Wolcott2018-zb} use bidirectional LSTM on character embeddings followed by dense layers to produce record-level representations.
As the only method, \citeauthor{Wolcott2018-zb} actually go straight from characters to record representation.
\citet{Kasai2019-ii} use bidirectional GRU on word embeddings to get an attribute-level representation.
As the only surveyed method, \citet{Gottapu2016-ho} apply a simple convolutional layer to word embeddings.
Lastly, both DeepER \cite{Ebraheem2018-ws} and AutoBlock \cite{Zhang2019-or} have networks solely aimed at finding good representations for use in blocking (see Section~\ref{sec:blocking}).
DeepER uses bidirectional LSTM on word embeddings to get a record-level representation (but also show a simple averaging approach is competitive).
Somewhat differently,
AutoBlock applies bidirectional LSTM and self-attention on word embeddings to get an attribute representation
and represents records as a weighted sum of attributes.

\paragraph{Interdependent representation}
DeepMatcher \cite{Mudgal2018-cx} explores several ways of building attribute representation from word embeddings.
The one with the highest representational power, Hybrid, uses a combination of bidirectional GRU and decomposable attention \cite{Parikh2016-jx} across records.
Unique among the surveyed methods,
Seq2SeqMatcher \cite{Nie2019-tw} structures records as sequences of \texttt{(attribute, word)} pairs.
The embedding of such a pair is a concatenation of a custom embedding for the attribute and a fastText embedding for the word itself.
The record-level representation is produced through an attention technique between the sequences of two records.
\citet{Brunner2020-tn} treat a record pair as a sequence of attribute value sub-word tokens,
while Ditto \cite{Li2020-jp} model record pair as a sequence of alternating attribute name and value sub-word tokens.
Both let each token keep its own representation throughout the representation building layers.
These Transformer networks take interdependent representation to an extreme,
as each token depends on all other in every Transformer layer.
Finally, Hi-EM \cite{Zhao2019-yv} is the only method which uses all the four explicit representation levels.
It applies a combination of bidirectional LSTM, self-attention within the record, and attention across records — both from its standard character embeddings to word vectors and from its word vectors to its attribute vectors.
For the record-level representation, it simply concatenate the attribute vectors.

\subsection{Schema matching}

Given two data sources $A$ and $B$, we divide the ways in which the schemas can be related in three:
\begin{itemize}
    \item \textbf{Aligned schemas}:
    Both data sources use the same schema.
    In other words, $\forall_{i \in \{1, 2, ..., n\}} (A_i = B_i)$ and $n = m$.
    \item \textbf{Misaligned schemas}:
    Both data sources have the same attributes, but not in the same order.
    In other words, there exists a bijective relation $H \subset \{A_i\}_{i=1}^n \times \{B_j\}_{j=1}^m$ such that $\forall A_i B_j: ((A_i, B_j) \in H) \rightarrow (A_i = B_j)$.
    \item \textbf{Incompatible schemas}:
    There is no simple correspondence.
    In other words, there does not exist such a bijective relation as described above.
\end{itemize}
With aligned schemas,
there is no need for schema matching.
For misaligned schemas,
finding a one-to-one correspondence between attributes is sufficient,
while in the general case of incompatible schemas,
more complex connections must be uncovered.
Many schema matching techniques are concerned with the former case, misaligned schemas.
For entity matching,
the goal is usually to find out which attributes should be compared in the downstream task where records are compared;
we want to find the pairs $(A_i, B_j)$ of attributes that are semantically related.
So one attribute can be compared to several attributes from the other data source -- implying incompatible schemas.

An additional challenge that might occur is dirty attribute values -- values that should have been in another attribute \cite{Mudgal2018-cx, Nie2019-tw}.
In such cases,
we need to compare attributes that might not necessarily be semantically related in order to be robust to noise.

There are two sources of information when doing schema matching.
There are the actual attribute values in records,
and there is the attribute metadata.
Attribute metadata will often simply be a name (e.g., \texttt{title}, \texttt{author}, etc).

When it comes to neural networks for schema matching,
there are essentially four approaches in the surveyed methods:

\subsubsection{Learn attribute matching from clusters}
SEMINT \cite{Li1994-sd,Li2000-le}, SMDD \cite{You_Li2005-jw}, and NNSM \cite{Zhang2014-bu} specifically target schema matching.
They all first create training data by performing unsupervised clustering of attributes,
and then use that data to train a multilayered perceptron\footnote{A multilayered perceptron is a simple feedforward network using only fully connected layers.} (MLP).
SEMINT uses a Self-Organizing Map \cite{Kohonen1987-zd} to cluster the feature vectors of attributes in data source $A$ into categories.
The attribute features are handcrafted and are based on both schema metadata and attribute values.
The category clusters are then used as labeled data to train an MLP with one hidden layer.
Given an attribute feature vector,
the network scores its similarity to these cluster categories,
and this is used to match the attributes of data source $B$ to the categories of $A$.
SMDD follows a similar strategy, but uses the distribution of attribute values and a hierarchical clustering technique.
Somewhat differently, NNSM clusters pairs of attributes into either being similar or dissimilar based on similarity scores of four traditional matchers.
Next, they train an MLP with two hidden layers to classify a pair of attributes as either similar or dissimilar using the clusters as training data.

An interesting aspect of these schema matching methods is their lack of need for human-labeled training data.
The methods that learn from clusters generate training data by using more traditional unsupervised manual methods.
As \citet{Zhang2014-bu} explains it,
they are essentially using neural networks as a way to combine several traditional methods.

\subsubsection{Learn schema mapping}
\citet{Nin2006-ld} translate records from source $A$ to records following the schema of $B$.
They train a network for each attribute in $B$,
which can translate a record $a \in A$ to a record of $B$'s schema.
Working only on purely numeric data, they are able to simply use records from $A$ as input and output values as the translated record from a neural network.
The network effectively transforms incompatible schemas into aligned schemas.
This approach can resolve the schema matching problem for downstream tasks but also does a lot of the heavy lifting of the record pair comparison by attempting to project records from $A$ to corresponding records in $B$.

\subsubsection{Compare attribute representations}
\label{sec:schema-matching-attribute-embeddings}
\citet{Nozaki2019-sg} do schema matching by thresholding the cosine distance between attribute vectors.
The attribute vectors are found by first summing up the pretrained word embeddings for each attribute in each record
and then simply averaging per attribute across all records.
Even though it relies on pretrained word embeddings,
the method itself is unsupervised.
The distance threshold was simply found experimentally.

\subsubsection{Learn jointly with comparison and classification}
While schema matching has traditionally been dealt with as a separate task,
as with the methods above,
\citet{Nie2019-tw}, \citet{Brunner2020-tn}, and \citet{Li2020-jp} incorporate it as part of their deep learning approach for comparing and classifying record pairs.

As explained in the previous subsection, Seq2SeqMatcher \cite{Nie2019-tw} structure records as sequences of \texttt{(attribute, word)} tokens,
and then solve entity matching as a sequence matching problem.
The embedding of such a token is a concatenation of a custom embedding for the attribute and a fastText embedding for the word itself.
Notice that no attribute metadata is used.
Treating the input in this way enables the neural network to learn how to compare values across attributes.
Specifically, the authors use a bidirectional attention mechanism between token embeddings from two records,
and then use only the max $k$ attention scores to get the soft-attended representation of a token.
Using only the $k$ largest attention scores, effectively setting the rest to zero, helps the model compare only relevant tokens and ignore irrelevant tokens.

\citet{Brunner2020-tn} preserve no information about the attributes other than the order (which, of course, may differ across schemas), and simply treat a record as a sequence of sub-word tokens.
They rely entirely on the powerful attention mechanism in the Transformer network to do the schema matching using positional information provided through the input and whatever insight and correlation the attribute values provide.
Ditto \cite{Li2020-jp}, on the other hand, explicitly incorporate attribute name sub-word tokens in the input sequence,
which gives more information to the Transformer network to perform schema matching.
In contrast to Seq2SeqMatcher, Ditto uses the attribute name instead of a randomly initialized embedding
--- enabling the network to exploit knowledge from its language modeling pretraining that the attribute name might signal.

\subsection{Blocking}
\label{sec:blocking}

Few methods try to use neural networks for blocking, as seen in Table~\ref{fig:process}.
The only two methods, DeepER \cite{Ebraheem2018-ws} and AutoBlock \cite{Zhang2019-or},
embed records into a high-dimensional metric space and then do nearest neighbor search to filter down the cartesian product $A \times B$ to a candidate pair set $C$.
They both use cosine distance as a metric,
and the networks are implicitly trained to produce record representations close to each other for matching records and far from each other for nonmatching records.
Finding the nearest neighbors in high-dimensional spaces is computationally infeasible,
so to make it more feasible,
they perform approximate nearest neighbor search.
Then there is no guarantee to find the nearest neighbors, but rather a high probability.
Both methods do this using locality sensitive hashing (LSH) \cite{Indyk1998-tv}, which is a well-studied technique \cite{Wang2014-ru}.

The two methods follow the same high-level strategy, but they have some important differences.
The networks themselves that are responsible for building a good record representation are differentiated in Section~\ref{sec:data-preprocessing}.
DeepER trains its network end-to-end with comparison and classification of record pairs.
The record representations are compared using either elementwise subtraction or multiplication,
and then a dense layer performs the classification.
AutoBlock, in comparison, trains specifically for blocking with a custom loss applied directly to the cosine distance between records.
For the actual nearest neighbor pruning, they use two different LSH methods.
DeepER uses hyperplane LSH \cite{Charikar2002-rb,Jain2010-ex}, a well-studied method that is known to be easy to implement and often fast in practice.
AutoBlock uses cross-polytope LSH \cite{Andoni2015-mg}, which has the benefit of theoretically optimal query running time while also being efficient in practice.
Both use multiprobing with distance 1.

\subsection{Record pair comparison}
\label{sec:comparison}

Central to matching is to assess the similarity of two records, both syntactically and semantically.
The surveyed neural networks will generally produce some distributed representation at either attribute- or record-level and then compare the representations.
We consider the layers in a network that are responsible for reducing from representations per record to representations across records appropriate for classification as record pair comparison layers.
Or, to put it simply, those layers producing the similarity vector $S$ from per-record representations.
We will look at three central characteristics of how comparison is performed:

\begin{table}
    \centering
    \caption{
        Overview of how the surveyed methods perform record pair comparison.
        $^+$Other options were tried, but this was found to be preferential.
        $^*$The most expressive model (BiLSTM-based) does non-attribute-aligned comparison, while the simpler averaging model is attribute-aligned.
    }
    \small
    \begin{tabular}{lM{22mm}M{22mm}M{26mm}}
        \toprule
        Method & Attribute-aligned & Distributed similarity & Attention-based \\ \midrule%\hline
        \citet{Kooli2018-sr}                 & No  & Yes       & No          \\ % \hline
        DeepMatcher \cite{Mudgal2018-cx}     & Yes & Yes$^+$   & Words$^+$   \\ % \hline
        \citet{Wolcott2018-zb}               & No  & No        & No          \\ % \hline
        DeepER \cite{Ebraheem2018-ws}        & No$^*$ & Yes    & No       \\ % \hline
        MPM \cite{Fu2019-ko}                 & Yes & Both      & (Partially) words \\ % \hline
        \citet{Kasai2019-ii}                 & Yes & Yes       & No          \\ % \hline
        Seq2SeqMatcher \cite{Nie2019-tw}     & No  & Yes       & Words       \\ % \hline
        Hi-EM \cite{Zhao2019-yv}             & Yes & Yes       & Characters, words \\
        \citet{Brunner2020-tn}               & No  & Yes       & Character N-grams \\
        Ditto \cite{Li2020-jp}               & No  & Yes       & Character N-grams \\
        \bottomrule
    \end{tabular}
    \label{tab:comparison}
\end{table}

\subsubsection{Attribute-aligned comparison}
If one assumes the two data sources $A$ and $B$ to have aligned schemas,
one can compare attributes in a one-to-one fashion.
We then say the comparison is attribute-aligned.
The alternative is to perform comparison at record level,
as one will be less dependent on the schemas to be aligned.
The second column in Table~\ref{tab:comparison} shows which of the surveyed methods do attribute-aligned comparison.
DeepMatcher \cite{Mudgal2018-cx} and \citet{Kasai2019-ii} both compare attributes one-to-one before they combine the similarity representation to record level.
To handle cases where non-attribute-aligned comparison is necessary because the data is dirty and are partially put in wrong attributes,
DeepMatcher merges all attributes to one long string attribute -- essentially reducing the problem to a string matching problem.
This is, of course, something most attribute-aligned methods can do to overcome this restriction,
but then the information carried by the attribute separation is lost.
Hi-EM \cite{Zhao2019-yv} does actually not align the comparison of attribute representations,
but the attribute representations have been produced by implicitly comparing characters and words through an attention mechanism across aligned attributes.

\subsubsection{Distributed similarity}
When two distributed representations are compared, one can either produce another distributed representation for the similarity of them
or reduce the representations down to a nondistributed similarity representation -- usually a scalar.
The third column in Table~\ref{tab:comparison} shows which of the surveyed methods make distributed similarity representations.
As can be seen, the majority of the surveyed methods do.
Typical ways of computing these similarities include vector difference, Hadamard product, or concatenation.
The only with nondistributed similarities, \citet{Wolcott2018-zb},
use cosine distance to compute the similarity.

Nondistributed similarities have the benefit of reducing complexity and training time,
but at the cost of expressiveness.
The increased expressive power of distributed similarities has to be matched by a classifier able to use it.
\citet{Mudgal2018-cx} reported that distributed similarities outperform nondistributed.
In addition, they found vector difference to be significantly better than concatenation when used after representation layers that do not use cross-attention.
MPM \cite{Fu2019-ko} stands out since it combines both distributed and nondistributed similarity.
It uses multiple classical similarity measures and two versions of DeepMatcher \cite{Mudgal2018-cx} in parallel and let the network effectively choose a similarity representation through a softmax.

\subsubsection{Cross-record attention}
As we saw in Section~\ref{sec:representation-levels},
some methods build distributed representations that are dependent on the record to be compared to through attention mechanisms.
They are essentially peeking at what is to come, which enables them to focus on what is important for the comparison.
The fourth column in Table~\ref{tab:comparison} summarizes which methods use cross-attention and at which representation levels.
DeepMatcher \cite{Mudgal2018-cx} uses attention between words across records,
while Hi-EM \cite{Zhao2019-yv} uses attention between both character- and word-level representations across records.
Both restrict their attention mechanism within each attribute, since the comparison is attribute-aligned.
In contrast, Seq2SeqMatcher \cite{Nie2019-tw}, \citet{Brunner2020-tn}, and Ditto \cite{Li2020-jp}, are able to use attention between all sub-words/words across the compared records.
The Transformer networks take cross-attention all the way by relying almost exclusively on attention throughout the architecture and not making any distinction between self-attention and cross-attention.

\subsection{Classification}

\begin{table}
    \centering
    \caption{
        Overview of which neural network layers the surveyed methods use for classification.
        $^*$The authors emphasize that any machine learning classifier can be used and do not explicitly favorize a neural network.
    }
    \small
    \begin{tabular}{lm{80mm}}
        \toprule
        Method & Classification layers \\
        \midrule %\hline
        \citet{Pixton2006-ko}            & Two linear layers with custom sparsity pattern, threshold on ratio of match and mismatch score \\ % \hline
        \citet{Wilson2011-qe}            & Single perceptron, threshold   \\ % \hline
        \citet{Tran2014-ww}              & MLP, softmax                  \\ % \hline
        \citet{Gottapu2016-ho}           & Linear layer, softmax \\ % \hline
        \citet{Reyes-Galaviz2017-ek}     & Two-layered MLP with custom sparsity pattern, threshold \\ % \hline
        \citet{Kooli2018-sr}             & MLP, LSTM, or CNN \\ % \hline
        DeepMatcher \cite{Mudgal2018-cx} & Two-layered MLP with Highway-connections \cite{Srivastava2015-hm}, softmax \\ % \hline
        DeepER \cite{Ebraheem2018-ws}    & Single dense layer$^*$ \\ % \hline
        MPM \cite{Fu2019-ko}             & Two-layered MLP with Highway-connections \cite{Srivastava2015-hm}, softmax \\ % \hline
        \citet{Kasai2019-ii}             & Two-layered MLP with Highway-connections \cite{Srivastava2015-hm}, softmax \\ % \hline
        Seq2SeqMatcher \cite{Nie2019-tw} & Two-layered MLP, softmax \\ % \hline
        Hi-EM \cite{Zhao2019-yv}         & Single dense layer \\
        \citet{Brunner2020-tn}           & Single dense layer, softmax \\
        Ditto \cite{Li2020-jp}          & Single dense layer, softmax \\
        \bottomrule
    \end{tabular}
    \label{tab:classification}
\end{table}

Compared to the other steps,
there is less variance in how the surveyed methods perform classification.
Generally,
they take in a similarity vector $S$ and do binary classification.
Deeming a record pair matching or not.
As an exception to this approach,
\citet{Gottapu2016-ho} classify records $a \in A$ directly to a corresponding record $b \in B$,
treating matching the problem as a multiclass classification with $|B|$ classes.

$S$ can be from a separate procedure, like string similarity measures, or upstream layers in the same neural network.
The first was more common in earlier methods,
while the former is common among newer deep learning methods.
Nonetheless,
the actual networks or layers used for classification are relatively similar.
Table~\ref{tab:classification} shows how each method's classification network is built -- either standalone or as the final layers of a larger network.
We see that most are variations of the same theme of MLP with softmax at the end.

\section{Contributions from deep learning}
\label{sec:contributions-from-deep-learning}

In the previous section we dissected the use of neural networks in all the surveyed methods using our process reference model.
We now take a step back and summarize which contributions deep learning provide entity matching.
Initially when neural networks were applied for entity matching, they were used merely as a classifier over feature vectors,
either for schema matching or determining if pair of records matched or not.
In the past few years, following the rise of deep learning,
we have seen not only an increase in the use of neural networks,
but also a broadening of the role they play in the entity matching process.

\subsection{Learned feature extraction and comparison}

Traditionally, as part of the preprocessing in entity matching,
it is common to transform the data through a range of handcrafted procedures to a format more suitable for comparison.
Important features are made more prominent and accessible to the steps downstream,
while less important features are filtered away.
Examples include phonetic encoding, removing punctuation, stemming, and expanding common domain-specific abbreviations.
The problem is that these procedures are highly dependent on the data sources,
and an expert has to decide which features are essential and how to extract them.
Such customization makes it harder to scale a solution across data sources and use cases.

In contrast, driven by the advances of deep learning for natural language processing,
more recent methods are able to learn feature extraction from less preprocessed records.
As mentioned in Section~\ref{sec:embeddings}, embeddings are used as a powerful gateway to letting neural networks work with text,
while hierarchical representation learning and sequence models make it possible to make semantically rich distributed representations of attribute values and records.
This alleviates the need for much of the ad-hoc handcrafted feature extraction procedures,
leaving mostly standardized feature extraction steps such as simple tokenization.
This is in line with the development in other domains using deep learning, where deep neural networks have been increasingly able to replace complex handtuned feature engineering \cite{LeCun2015-jz}.
Of course, it does not replace all forms of preprocessing when data sources in different formats and of different origins are to be matched.
One might, for example, need to extract records from some nonstandard XML or JSON format.
While current deep learning methods do not remove the need for such handcrafted ad-hoc preprocessing,
feature extraction can still be a very labor-intensive part of the preprocessing,
and so deep learning has the potential to lighten the manual load drastically.

In addition to feature extraction,
a point of significant handcrafted complexity is that of string similarity measures.
Traditionally,
one would use several string similarity measures between two records to produce a similarity feature vector
and then use a neural network, some other classification model, or rules to classify the pair as matching or nonmatching based on this similarity vector.
Doing comparison together with feature extraction in a deep neural network,
we are able to effectively learn how to do comparison of records.
The network will learn to extract features that are suitable for comparing records,
either for static comparison or for downstream layers in the network.
In cases where cross-attention is used,
the network is even able to learn to extract features tailored to be suitable for comparison against a specific record.
Compared to traditional approaches,
learning comparison in this way can remove the need for these complex handcrafted similarity measures.
Also,
since deep networks can produce powerful distributed representations,
this approach opens up a straightforward way to perform semantic comparison.
This means one can depend less on syntactic similarity, which is hard to do with handcrafted methods.

Transfer learning has been a catalyst to being able to train more powerful deep learning models with fewer labeled examples for the task at hand in both natural language processing \cite{Ruder2019-dz} and computer vision (where starting out with large networks \cite[e.g.,][]{He2015-iu} pretrained on enormous data sets like ImageNet \cite{Deng2009-pw} is common).
Deep learning methods for entity matching have also started to use such opportunities.
Using some of the readily available pretrained word embeddings is popular,
and is used as an efficient way of transfer learning from large general corpora.
\citet{Thirumuruganathan2018-po} outline different ways to transfer learn between entity matching task using weigted sums of word embeddings as distributed record representations.
Some methods also use transfer learning beyond word embeddings and pretrain models specifically for entity matching \cite{Kasai2019-ii, Zhao2019-yv}.
Furthermore, most recently, we have also seen the use of large, powerful, general pretrained language models that are fine-tuned to entity matching \cite{Brunner2020-tn,Li2020-jp}.
Compare this to traditional methods, which do not typically incorporate transfer learning.
While one can argue that this is to be expected, since deep learning methods will usually require more training examples,
the use of deep networks makes it possible to realize very powerful transfer learning scenarios,
which are hard to pull off with traditional methods.

\subsection{Coalescing the entity matching process}

\begin{figure}
    \centering
    \includegraphics{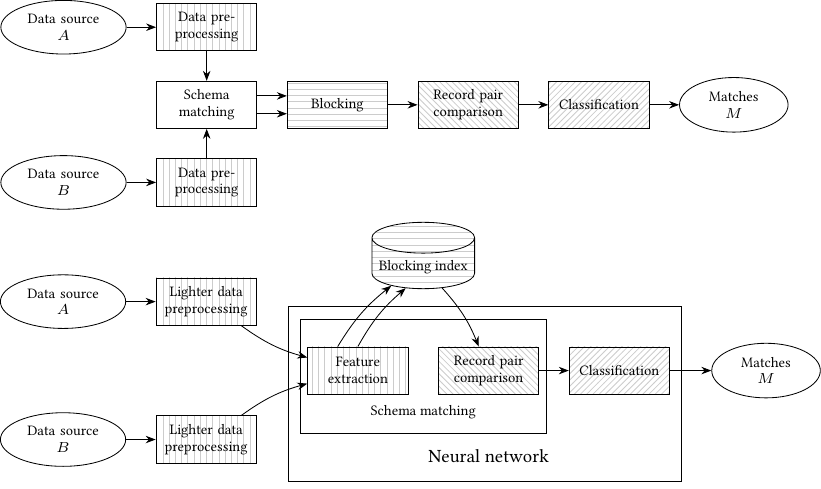}
    \caption{
        Illustration of the reference model for a deep learning entity matching process (bottom) together with the reference model for the traditional entity matching process (top)
        and how they relate to each other.
    }
    \label{fig:deep-learning-process}
\end{figure}

In Section~\ref{sec:process},
we introduced the reference model of the traditional entity matching process.
To support our discussion, we introduce a reference model for a deep learning-based entity matching process.
Figure~\ref{fig:deep-learning-process} depicts this reference model together with the traditional variant while highlighting how steps correspond between them.
One should note that none of the surveyed deep learning methods actually follow this exact process.
It is the essence of all the methods summed up in one view.

What all surveyed deep learning methods have in common is that they have fused together feature extraction, record pair comparison, and classification in a single step as a neural network.
The need for data processing as a separate step still exists, but with less focus on feature extraction.
In some ways, this development comes naturally.
The steps are carried out in more or less the same order, and one can (to a certain degree) distinguish between them in the neural network architecture.
But as we saw in Section~\ref{sec:process},
some methods have also explored using neural networks for schema matching and blocking.

\citet{Nie2019-tw} is the first deep learning method to incorporate schema matching in an end-to-end fashion together with record pair comparison and classification.
They essentially construct the feature extraction and record pair comparison part of the network in such a manner that it is possible to learn schema matching --
reducing the step to a built-in property of the network --
in effect, jointly training a model for both matching schemas and records.
Compared to solving schema matching as a separate task upfront,
this has the potential benefit of being able to adapt how the schemas are matched to how they are used downstream.
Other deep learning methods assume the schema to be the same for both data sources,
but to what degree and how this assumption is manifested in the method differ (see Section~\ref{sec:taxonomy}).

The surveyed methods tackling blocking with deep learning \cite{Ebraheem2018-ws,Zhang2019-or} rely on making a distributed representation of a record with a neural network,
and then finding the candidate pairs with approximate nearest neighbor search.
In other words, the network produces indexable feature vectors,
and through the use of a suitable index,
we find similar records in subquadratic time.
While it does not remove the blocking step,
it does reduce the core blocking mechanism to an indexing problem over some standard metric.
The network will learn how to do blocking by learning how to represent similar records close to each other in the metric space,
while the nearest neighbor search will always remain the same.
If one uses the same distributed representations downstream for record pair comparison, it also serves as a good way to align how blocking and record pair comparison evaluate similarity.
It is not unusual in traditional methods that these two steps, who both at their core do record comparison, have two considerably different ways of measuring similarity.

In computer vision and natural language processing,
one has traditionally operated on large, complex, and heavily engineered pipelines.
They are often divided into distinct steps that have been worked on separately.
In contrast,
deep learning methods in these fields have evolved to do what previously was done in several distinct steps in one go with a deep network \cite[e.g.,][]{Sutskever2014-ya, Ren2015-it},
making it possible to train on tasks end-to-end.
In a similar fashion, we can see deep learning increasingly reshaping the entity matching process by coalescing traditional steps.
This has at least two immediate benefits.
First, it effectively reduces the number of steps in the process,
and second, it makes it possible to train an increasing part of the process end-to-end.
The main enabling factor is the representation learning in modern deep learning techniques for natural language processing,
which is able to tie all of these together.
As we have seen, powerful feature extraction can remove much of the need for manual feature engineering in schema matching, blocking, and record pair comparison.
When in addition these steps are able to share the feature extraction and become part of an end-to-end network covering a large portion of the entity matching process,
it can be translated to a potentially great reduction in complexity when building entity matching pipelines.
Because the complexity of the process lies not only in the individual steps, but also in the interaction between them.
Tuning and getting multiple heavily engineered steps to work smoothly together can be difficult.

\section{Taxonomy of deep neural networks for entity matching}
\label{sec:taxonomy}

\begin{figure}
    \centering
    \includegraphics{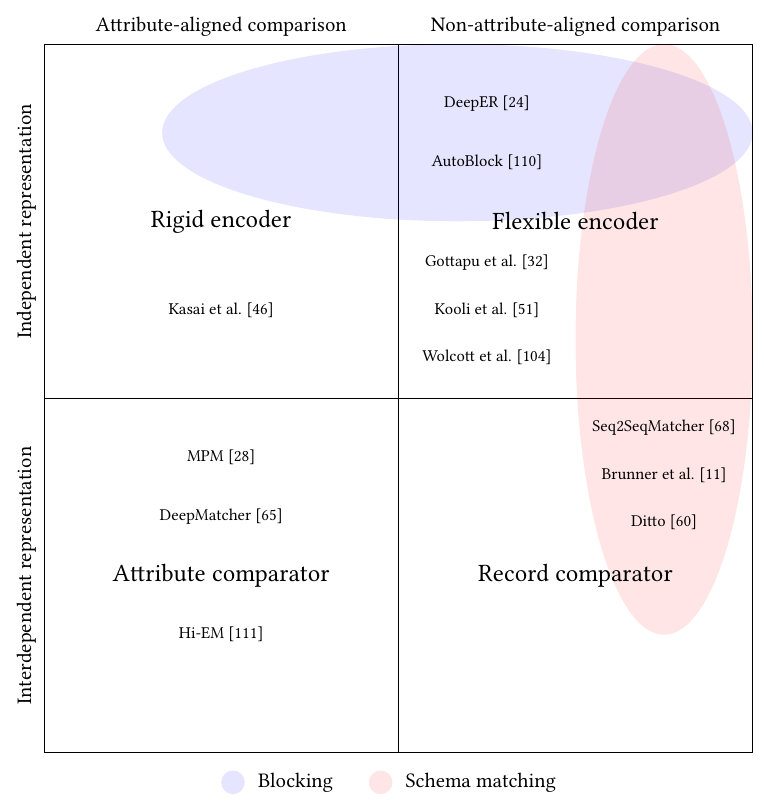}
    \caption{
        Taxonomy of deep neural networks for entity matching.
        Split into four categories along two axes that represent two binary properties.
        Methods within the blue and red colored areas are methods that address blocking and schema matching, respectively.
    }
    \label{fig:taxonomy}
\end{figure}

\begin{figure}
    \centering
    \includegraphics{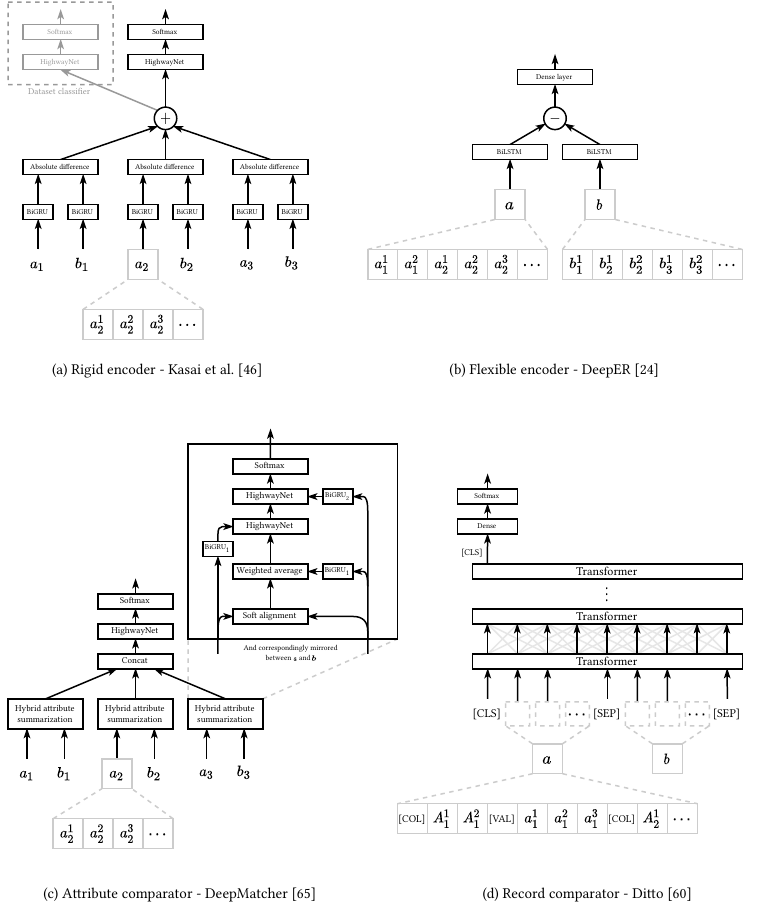}
    \caption{
        High-level illustration of four deep learning architectures --- one for each category in the taxonomy.
        Note that the input is assumed to have already been replaced with its embedding.
    }
    \label{fig:architectures}
\end{figure}

We have seen how neural networks serve different purposes in the entity matching process and vary in how they do so.
Deep learning has led to an increase in the use of neural networks for entity matching in the past few years,
making deep networks the norm.
All of these deep learning methods have in common that they reduce the need for tedious handcrafting of features, but how the networks are structured at a high level differs in important ways, which impact their ability to interact with other steps in the entity matching process.
The schema matching and blocking steps are especially interesting in this regard,
since they have traditionally been solved with specialized methods separate from record pair comparison and classification.
With this in mind,
we propose a taxonomy of deep neural networks for entity matching consisting of four categories.
There is no concrete definition of what constitutes a \textit{deep} neural network,
so for the purpose of this taxonomy we only consider networks that do feature extraction\footnote{Note we also exclude \citet{Nozaki2019-sg}, since the method do not represent a neural network itself.} (basically those who are marked for the data preprocessing step in Table~\ref{tab:method-process-matrix}).
The categories are decided by two binary properties that were introduced in Section~\ref{sec:process}:
1) Whether comparison is attribute-aligned or not, and 2) whether representations used for comparison are independent or interdependent.
Figure~\ref{fig:taxonomy} shows the taxonomy and which category each surveyed deep neural network falls into.
In addition,
it highlights how schema matching and blocking are related to the categories and which of the methods leverage their neural networks to tackle these steps.
The two properties do not only reflect the high-level structural properties of the network,
but also the assumptions of the problem to be solved.
To help illustrate the taxonomy, we also show one representative deep learning network architecture for each category in Figure~\ref{fig:architectures}.

\subsection{Attribute-aligned or non-attribute-aligned comparison}
If the network is structured in a way that assumes records from the two data sources to have aligned schemas and constrain record comparison to one-to-one on attributes,
we say it has attribute-aligned comparison.
If it does not, we say it has non-attribute-aligned comparison.
Notice that comparison does not necessarily need to be explicit record pair comparison,
but can also be implicit comparison through, for example, indexing of distributed representations.

Attribute-aligned comparison is a powerful way to incorporate prior knowledge into a method,
potentially making training easier and more efficient.
At the same time,
it prevents the possibility of performing schema matching.
Non-attribute-aligned comparison does not necessarily imply the method is being used for schema matching,
but that it is in its nature more compatible and should be easier to adapt to performing it.

\subsection{Independent or interdependent representation}

If the network relies on seeing a pair of records that are to be compared to produce representations of the records,
we say it generates interdependent representations.
Otherwise, we say it generates independent representations.

Interdependent representations have the benefit of being able to observe what they are being compared to.
Intuitively, it is easier to compare two things if we can look at both at the same time.
If we are only allowed to look at one thing at a time,
then it is harder to know what to focus on.
On the other side,
by being dependent on seeing the record to be compared to produce a representation,
one has essentially made it impossible to compare a large number of records in any subquadratic way.
For independent representations,
it is possible to generate a representation of each record once and implicitly compare them through a form of indexing,
effectively opening up the possibility of doing blocking with the representations made from the network.

\subsection{The four categories}
\begin{itemize}
    \item \textbf{Rigid encoders} are networks that produce independent attribute representations and then perform comparison.
    Even though none of the surveyed methods do,
    it is possible to use such independent attribute representations for blocking purposes.
    \item \textbf{Flexible encoders} are networks that produce independent record representations and then perform comparison.
    It is possible to perform blocking using these independent representations,
    as shown by DeepER \cite{Ebraheem2018-ws} and AutoBlock \cite{Zhang2019-or}.
    If the representations can be built from records with two different schemas,
    one can also effectively perform schema matching,
    but none of the current methods do.
    \item \textbf{Attribute comparators} are networks that use cross-attention only within the same attribute.
    They are inherently focused on doing aligned attribute-to-attribute comparison.
    DeepMatcher \cite{Mudgal2018-cx} has attribute-to-attribute comparison explicitly designed into the network architecture.
    Hi-EM \cite{Zhao2019-yv} takes it a step further by training individual networks per attribute.
    There is no easy way to perform blocking or schema matching with such a network architecture.
    \item \textbf{Record comparators} are networks that use cross-attention at record level without being constrained by the boundaries of attributes.
    They can learn how to compare across attributes,
    making it possible to overcome misaligned schemas and even incompatible schemas, depending on the network structure.
    Seq2SeqMatcher \cite{Nie2019-tw} was the first of the surveyed method in this category,
    and it is able to handle incompatible schemas.
    Transformer-based networks, which can naturally work as record comparators, have later shown to provide state-of-the-art performance \cite{Brunner2020-tn,Li2020-jp}.
\end{itemize}

\section{Evaluation}
\label{sec:evaluation}

\subsection{Metrics}
An entity matching system can be evaluated in several ways.
Since it is fundamentally a very skewed binary classification problem with few positives compared to negatives,
it is natural to use precision/recall measures -- popular metrics from information retrieval and machine learning classification.
And while precision and recall (as well as accuracy) are sometimes reported,
the most prominent is to report and evaluate the matches using the $F_1$ measure.
Of course, while simple, this metric only measure one aspect of an entity matching system.
Different authors have therefore used additional metrics.
\citet{Kasai2019-ii} focus on the important issue of how many training examples a model needs,
and measure $F_1$ for different amounts of provided training examples.
Focused on scalability,
\citet{Wolcott2018-zb} report wall-clock running time for different sizes of the data sources.

It is also possible to evaluate intermediate steps in isolation in addition to the end-to-end result.
Some methods specifically target blocking,
and so specifically measure the outcome of that step.
One could evaluate blocking the same way as the end result,
using $F_1$.
But remember that for blocking we are mainly interesting in getting high recall,
and less interesting in precision as long as the number of candidate pairs $|C|$ is sufficiently lower than the size of the Cartesian product $|A \times B|$.
So the surveyed methods report both recall and some variant of reduction ratio from data sources to candidate set.
\citet{Ebraheem2018-ws} report $RR = \frac{|C|}{|A \times B|}$,
while \citet{Zhang2019-or} report $P/E = \frac{|C|}{|A|}$.

\subsection{Datasets}

The early approaches to entity matching were mostly concerned with matching personal records (census data or medical records).
Such datasets are usually not publicly available due to privacy concerns.
Today,
methods are evaluated on data from different domains.
A substantial of amount of reported results are from closed datasets,
but we're seeing increasingly more open datasets being used.
Some of the most popular open datasets include domains such as publications, restaurants, products, songs, and companies \cite{Das2016-hs, noauthor_undated-ot, Kopcke2010-ck}.
See Table~\ref{tab:public-datasets} for an overview of the most popular public datasets among the surveyed methods.
In order to test certain scenarios,
some authors also artificially construct new datasets based on existing ones,
either by reducing the data quality \cite[e.g.,][]{Mudgal2018-cx} or constructing new records \cite[e.g.,][]{Ioannou2013-fi}.

\begin{table}
    \centering
    \caption{
    Overview of public datasets that have been used by at least two of the surveyed publications.
    \# Records lists the number of records in each data source,
    \# Matches denotes the number of matches between them,
    and \# Pos / \# Candidates denotes the number of positive examples among all the record pair candidates in an agreed-upon subset of the potential matches.
    The latter is used when blocking is not part of the experiment.
    }
    \small
    \begin{tabular}{llcccM{17mm}}
        \toprule
        Dataset           & Domain      & \# Records    & \# Attributes  & \# Matches & \# Pos / \newline \# Candidates   \\
        \midrule
        Abt-Buy \cite{Kopcke2010-ck,Das2016-hs}             & Product     & 1081 - 1092   & 3              & 1096       & 1028/9575  \\ % 3
        Amazon-Google \cite{Kopcke2010-ck,Das2016-hs}       & Software    & 1363 - 3226   & 3              & 1300       & 1167/11460 \\ % 3
        Beer \cite{Das2016-hs}                              & Beer        & 3274 - 4345   & 4              &            & 68/450     \\ % 2
        DBLP-ACM \cite{Kopcke2010-ck,Das2016-hs}            & Citation    & 2616 - 2294   & 4              & 2224       & 2220/12363 \\ % 6
        DBLP-Scholar \cite{Kopcke2010-ck,Das2016-hs}        & Citation    & 2616 - 64263  & 4              & 5347       & 5347/28707 \\ % 6
        Fodors-Zagats \cite{noauthor_undated-ot,Das2016-hs} & Restaurant  & 533 - 331     & 6              & 112        & 110/946    \\ % 3
        iTunes-Amazon \cite{Das2016-hs}                     & Music       & 4875 - 5619   & 8              &            & 132/539    \\ % 3
        Walmart-Amazon \cite{Das2016-hs}                    & Electronics & 2554 - 22074  & 5              & 1154       & 962/10242  \\ % 5
        % DBLP-Citesheer    & Citation    & & & & \\ % 1
        % Cora              & Citation    & & & & \\ % 1
        % Zomato-Yelp       & Restaurant  & & & & \\ % 1
        \bottomrule
    \end{tabular}
    \label{tab:public-datasets}
\end{table}

\subsection{Experimental results}

\subsubsection{Comparing surveyed methods}

It has historically been hard to directly compare reported experimental results.
Either because they do not target and measure the same aspect of the entity matching process, do not use the same data, or do not evaluate in the same way.
For example, several methods have been evaluated on some of the same datasets but used completely different train-test splits (both the actual selection and train-test ratio).
There is a general lack of widespread and agreed upon benchmarks that would make comparison across many methods straightforward --- like we have seen in core computer vision and NLP tasks.
To a large degree, on has relied on authors to reimplement and run other methods within their own evaluation setup.
So for the majority of the surveyed methods, we can not make any comparison using reported experimental data.

\begin{figure}
    \centering
    \includegraphics{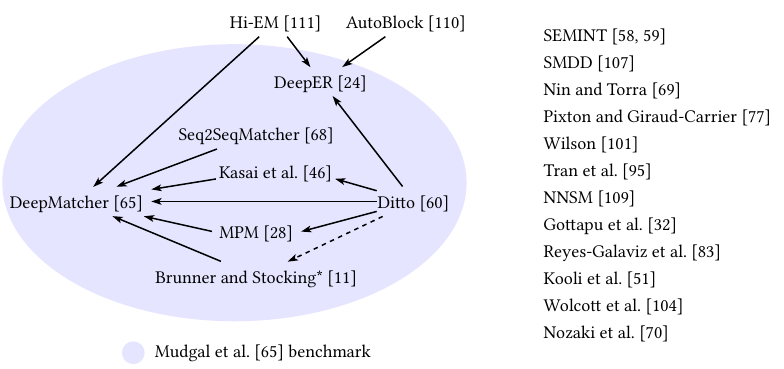}
    \caption{
        Overview of which of the surveyed methods has been compared experimentally to each other and which have at least partially been subject to the same benchmarks.
        Arrows indicate that the method at the head was compared to the method at the tail in the latter's publication.
        Note that they are not necessarily transitive, and they do not all test the same task (i.e., some compare blocking, others only matching after blocking).
        The blue area is methods that have been tested on at least a subset of the public benchmark from \citet{Mudgal2018-cx}.
        $^*$As pointed out by \citet{Li2020-jp}, the authors do model selection using the test set --- effectively leaking from the test set and making the results slightly unfit for comparison with others.
    }
    \label{fig:compare-experiments}
\end{figure}

It is, however, possible to partially compare some of the more recent methods.
Figure~\ref{fig:compare-experiments} shows which surveyed methods that have experimental results that let you compare it to other methods.
Mainly due to the adaptation of some of the experiments done by \citet{Mudgal2018-cx} as a benchmark,
but also due to authors of a method running new experiments on earlier methods in their evaluation setup.
The \citet{Mudgal2018-cx} benchmark let us compare (at least partially) several methods at once.
Table~\ref{tab:mudgal-benchmark} presents an overview of all available experimental results on those specific benchmark experiments.
We observe that the Transformer-based networks, especially Ditto \cite{Li2020-jp}, seems to be the current state-of-the-art.
Ditto does use some domain-specific optimizations, but the authors report strong results for a baseline without those optimizations.
Note that this benchmark only addresses certain aspects of the entity matching process.
Comparable results for aspects such as training time, prediction latency, and label efficiency can not be collected in the same way for several methods --- 
even though some results and comparison between pairs of surveyed methods have been reported.
And neural methods addressing blocking, using active learning, or using transfer learning are not in a state where any significant overview can be provided.

\begin{table}[]
    \centering
    \caption{
        Overview of reported results from the surveyed methods on the public part of the benchmark from \citet{Mudgal2018-cx}.
        We have also included reported results \cite{Mudgal2018-cx} from Magellan (MG) as a state-of-the-art classical non-neural method.
        All experiments use the post-blocking record pair candidates described in Table~\ref{tab:public-datasets}.
        The \textit{structured} versions are the unaltered datasets where attributes are aligned,
        while in the \textit{dirty} variant other attributes are moved to the \texttt{title} attribute with a 50\% probability.
        For the \textit{textual} dataset, Abt-Buy, all attributes are long textual descriptions.
        $^+$We only report the Hybrid model, which is generally considered the strongest.
        $^\dagger$The results from \citet{Ebraheem2018-ws} are not comparable to the others due to different setup, so reproduced results from \cite{Li2020-jp} are used instead.
        Note that those results do not involve the blocking method from \cite{Ebraheem2018-ws}.
        $^*$Not entirely fit for comparison since the authors do model selection using the test set.
    }
    \scriptsize
    \begin{tabular}{lcccccccc}
        \toprule
        Dataset & MG \cite{Konda2016-mu} & DM$^+$ \cite{Mudgal2018-cx} & DeepER$^\dagger$ \cite{Ebraheem2018-ws} & MPM \cite{Fu2019-ko} & Kasai \cite{Kasai2019-ii} & S2SM \cite{Nie2019-tw} & B\&S$^*$ \cite{Brunner2020-tn} & Ditto \cite{Li2020-jp} \\
         \midrule
        \textbf{Structured} & &            &      &       &      &      &       \\
        Amazon-Google  & 49.1 & 69.3 & 56.08 & 70.7 &       &      &      & 75.58 \\
        Beer           & 78.8 & 72.7 & 50.00 &      &       &      &      & 94.37 \\
        DBLP-ACM       & 98.4 & 98.4 & 97.63 &      & 98.45 & 98.9 &      & 98.99 \\
        DBLP-Scholar   & 92.3 & 94.7 & 90.82 &      & 92.94 & 95.3 &      & 95.60 \\
        Fodors-Zagats  & 100.0& 100  & 97.67 &      &       &      &      & 100.0 \\
        iTunes-Amazon  & 91.2 & 88.0 & 72.46 &      &       &      &      & 97.06 \\
        Walmart-Amazon & 71.9 & 66.9 & 50.62 & 73.6 &       & 78.2 &      & 86.76 \\
        \midrule                  
        \textbf{Dirty} &      &       &      &      &       &      &              \\
        DBLP-ACM       & 91.9 & 98.1 & 89.62 &      &       & 98.4 & 98.9 & 99.03 \\
        DBLP-Scholar   & 82.5 & 93.8 & 86.07 &      &       & 94.1 & 95.6 & 95.75 \\
        iTunes-Amazon  & 46.8 & 74.5 & 67.80 &      &       &      & 94.2 & 95.65 \\
        Walmart-Amazon & 37.4 & 46.0 & 36.44 &      &       & 68.3 & 85.5 & 85.69 \\
        \midrule                         
        \textbf{Textual} &    &      &       &      &       &      &      &       \\
        Abt-Buy        & 43.6 & 62.8 & 42.99 &      &       &      & 90.9 & 89.33 \\
         \bottomrule
    \end{tabular}
    \label{tab:mudgal-benchmark}
\end{table}

\subsubsection{Deep learning vs. traditional methods}

Several of the surveyed methods have evaluated deep learning approaches against more traditional approaches \cite{Zhang2019-or, Nie2019-tw, Kasai2019-ii, Wolcott2018-zb, Kooli2018-sr, Mudgal2018-cx, Ebraheem2018-ws, Fu2019-ko, Brunner2020-tn}.
The results are generally promising for deep learning approaches,
but traditional methods are often competitive.
Most methods compare themselves to Magellan \cite{Konda2016-mu},
which is considered a state-of-the-art traditional entity matching system,
using $F_1$ scores.
\citet{Mudgal2018-cx} report that DeepMatcher mostly outperforms Magellan with a few exceptions (see parts of the result in Table~\ref{tab:mudgal-benchmark}).
The relative strength of DeepMatcher to the traditional method increases as the data quality decreases,
and the same is reported about Seq2SeqMatcher \cite{Nie2019-tw}.
AutoBlock \cite{Zhang2019-or} is found to be especially strong against traditional blocking techniques for dirty data.
This may suggest that deep learning approaches' strength is most clearly seen when the data is noisy and heterogeneous.

\citet{Kasai2019-ii} investigate the use of transfer learning and active learning for their deep learning model.
They find Magellan to significantly outperform their model when given few labeled training examples,
but they were able to make it substantially more competitive using transfer learning and active learning.
The model still performs favorably in comparison to the traditional methods when given enough examples.

\section{Future research}
\label{sec:challenges}

In Section~\ref{sec:contributions-from-deep-learning}, we saw which contributions deep learning have made to entity matching.
We will now take a look at both the challenges and opportunities for deep neural networks in entity matching,
both of which represent potential directions for future research.

\subsection{Challenges}

\subsubsection{Explainability and ease of debugging}
Entity matching, being a central data integration task,
is constructing data models that are consumed by people or machines for downstream tasks.
For many applications, it is crucial to trust the data source,
and being able to understand why something does not work is key.
Unfortunately, deep learning models are notoriously hard to interpret.
As steps in the entity matching process increasingly coalesce into a large neural network, as illustrated in Figure~\ref{fig:deep-learning-process},
we get fewer checkpoints along the way in the process that can easily be inspected.
We cannot see the output from each step in the same way anymore.
Therefore, figuring out why two records where matched or not matched is usually nontrivial.
There are a few techniques that are already used, such as looking at alignment scores,
but we are still far away from a comprehensive way of debugging neural networks for entity matching.

\subsubsection{Running time in interactive settings}
Human interaction is considered an important factor in entity matching \cite{Doan2017-pi}.
Users cannot generally be expected to wait for very long when they are supposed to do interactive work,
limiting the potential applications of deep learning for entity matching in interactive settings, since the running time for both training and prediction is high -- especially compared to cleverly engineered traditional pipelines with good blocking.

\subsubsection{Number of training examples}
As more steps in the entity matching process are addressed by deep learning-based techniques,
the more different steps rely on more training data.
Deep learning models are hungry for training examples,
and there is not always an abundance of them readily available.
Even while the use of deep learning can potentially reduce the need for manual labor in the form of feature engineering,
it might still be necessary to do expensive manual labeling.
While transfer learning and active learning could help,
there are not any pretrained entity matching models publicly available to train from.

\subsection{Opportunities}

\subsubsection{End-to-end approach with schema matching and blocking}
As we have seen,
deep neural networks are increasingly able to take over steps in the entity matching process.
But we have still not seen a method doing both schema matching and blocking together with record pair comparison and classification in an end-to-end solution with a neural network --
in other words,
an approach that implements the whole deep learning entity matching reference model in Figure~\ref{fig:deep-learning-process}.
The network would presumably be a \textit{flexible encoder},
belonging in the upper-right corner of our taxonomy.
It would have independent representations to make efficient blocking possible and non-attribute-aligned comparison to make schema matching possible.
This could be an important step toward tackling the entity matching problem in a streamlined end-to-end fashion.

\subsubsection{More open datasets}
One of the fundamental challenges when trying to develop an entity matching methods that will work across many datasets is the huge variation between domains and data sources.
While several open datasets are available,
we would like to see significantly more.
It is especially important to increase the diversity of the domains represented.
For example, many datasets are related to either research publications or consumer-focused products/services -- there are no industrial datasets.
This would not only enable more complete evaluations, but also provide data suitable for transfer learning.

\subsubsection{Standardized benchmarks}

It is not always easy to compare the different methods since they do not necessarily evaluate on the same data in the same way.
We have seen how standard open datasets with corresponding standard evaluation metrics have been a catalyst for advances in machine learning for fields such as computer vision \cite{Russakovsky2015-hh}, recommender systems \cite{Bennett2007-ox} and natural language processing \cite{Bojar2018-nz}.
We think there is great potential in similar agreed-upon benchmarks for entity matching.
This need not only be restricted to traditional precision/recall measures for the resulting matches.
It would also be of interest to standardize the evaluation of blocking techniques, transfer learning techniques, active learning techniques, and computational performance and efficiency.

\subsubsection{Publicly available pretrained models}
While there has been work on transfer learning \cite{Kasai2019-ii, Zhao2019-yv, Brunner2020-tn, Li2020-jp},
there are no pretrained models publicly available specifically for entity matching.
Having pretrained models to fine-tune could potentially speed up training and reduce the amount of necessary training examples.
This is challenging, because, as mentioned above, there is a huge variation across domains and data sources.
Each data source is different.
Building pretrained models that can be fine-tuned for a broad number of data sets and making them publicly available would be of huge benefit to the field.
This would, of course, benefit from the previous point about more open datasets.

An interesting approach for those networks with attribute-aligned comparison,
explored by \citet{Zhao2019-yv},
is to have pretrained models for different types of attributes (e.g., name, addresses, organization) in addition to generic ones.
Then one can mix and match models for each attribute one is about to match.
This, of course, only works for aligned schemas.
It can be even more interesting to look at networks with non-attribute-aligned comparison for such pretrained models,
since they can potentially offer pretrained schema matching,
which will be crucial to handle the large variety of data source schemas out there.
For this, Transformer-based networks have considerable potential since they are already built on top of heavily pretrained language models.

Pretrained models can belong to any of the four categories of our taxonomy.
Nonetheless, we find pretrained flexible encoders that can support both blocking and schema matching to be the most significant opportunity.
Specifically, because it would enable transfer learning jointly on the largest possible portion of the entity matching process.

\section{Conclusion}

We have seen how existing work has used neural networks in entity matching.
By using a reference model of the traditional entity matching process,
we have shown how the surveyed methods address different steps of the process and which techniques have been used at each step.

More recently, approaches based on deep learning techniques for natural language processing have emerged.
We looked at what such deep learning methods contribute to the entity matching task.
The central contribution is powerful hierarchical representation learning from text.
As we have seen,
this can alleviate most of the handcrafted feature engineering necessary in the data preprocessing,
which can ease the burden of manually tailored procedures in downstream steps such as record pair comparison.
Furthermore,
it is a driver in increasingly coalescing steps of the entity matching process into end-to-end neural networks performing several steps in one go,
effectively reducing the number of steps and enabling end-to-end training.
To give a clear view of how the entity matching process changes with such an increasingly coalesced deep learning approach, we propose a reference model for entity matching processes using deep learning.

To differentiate the deep neural networks used in the surveyed methods,
we introduced a taxonomy of deep neural networks for entity matching.
It focuses on two properties that are important in regard to how easy it is to support schema matching and blocking.

Lastly, we looked at potential directions for future research by discussing challenges and opportunities.
The challenges being explainability, running time in interactive settings, and the large need for training examples,
while for opportunities we think it would be interesting to develop a complete end-to-end approach with both schema matching and blocking,
exploring a new part of the deep neural network taxonomy.
We also see a lot of potential in trying to develop more open datasets, standardized benchmarks, and publicly available pretrained models for entity matching --- which have been important for other fields.

\begin{acks}
This work is supported by \grantsponsor{cognite}{Cognite}{https://cognite.com} and \grantsponsor{fr}{the Research Council of Norway}{https://www.forskningsradet.no} under Project \grantnum{fr}{298998}.
We thank the reviewers for valuable feedback and Carl F. Straumsheim for proofreading.
\end{acks}

\bibliographystyle{ACM-Reference-Format}
\bibliography{bib}


\end{document}